\PassOptionsToPackage{table,dvipsnames}{xcolor}
\documentclass[acmsmall]{acmart}
\AtBeginDocument{%
  }

\setcopyright{acmlicensed}
\copyrightyear{2025}
\acmYear{2025}
\acmDOI{XXXXXXX.XXXXXXX}

\acmJournal{PACMMOD}
\acmVolume{X}
\acmNumber{X (SIGMOD)}
\acmArticle{XX}
\acmMonth{02}

\usepackage{amsmath}
\usepackage{amsthm}
\usepackage{acronym}
\usepackage{graphicx}
\usepackage{booktabs}
\usepackage{multirow}
\usepackage{dsfont}
\usepackage{pdflscape}
\usepackage{algpseudocode}
\usepackage{algorithm}
\usepackage{makecell}
\usepackage{url}
\usepackage{balance}
\usepackage{tikz}
\usepackage{marginnote}

\definecolor{purple}{rgb}{0.88,0.88,1}

\theoremstyle{definition}

\newtheorem{fallacy}{Fallacy}
\newtheorem{theorem}{Theorem}
\newtheorem{corollary}{Corollary}

\newtheorem{example}{Example}

\newtheorem*{problem}{Problem}

\newcommand*\circled[1]{\tikz[baseline=(char.base)]{\node[shape=circle,draw,inner sep=1pt,fill=black,text=white] (char) {\footnotesize #1};}}

\newcommand{\rone}[1]{\textcolor{black}{#1}}
\newcommand{\rtwo}[1]{\textcolor{black}{#1}}
\newcommand{\rthree}[1]{\textcolor{black}{#1}}
\newcommand{\rometa}[1]{\textcolor{black}{#1}}
\newcommand{\rttmeta}[1]{\textcolor{black}{#1}}

\begin{document}

\title{Credible Intervals for Knowledge Graph Accuracy Estimation}

\author{Stefano Marchesin}
\email{stefano.marchesin@unipd.it}
\orcid{0000-0003-0362-5893}
\affiliation{%
  \institution{University of Padua}
  \city{Padua}
  \country{Italy}
}

\author{Gianmaria Silvello}
\email{gianmaria.silvello@unipd.it}
\orcid{0000-0003-4970-4554}
\affiliation{%
  \institution{University of Padua}
  \city{Padua}
  \country{Italy}
}

\renewcommand{\shortauthors}{Marchesin and Silvello}

\begin{abstract}
Knowledge Graphs (KGs) are widely used in data-driven applications and downstream tasks, such as virtual assistants, recommendation systems, and semantic search. The accuracy of KGs directly impacts the reliability of the inferred knowledge and outcomes. Therefore, assessing the accuracy of a KG is essential for ensuring the quality of facts used in these tasks. However, the large size of real-world KGs makes manual triple-by-triple annotation impractical, thereby requiring sampling strategies to provide accuracy estimates with statistical guarantees.
The current state-of-the-art approaches rely on Confidence Intervals (CIs), derived from frequentist statistics. 
While efficient, CIs have notable limitations and can lead to interpretation fallacies. 
In this paper, we propose to overcome the limitations of CIs by using \emph{Credible Intervals} (CrIs), which are grounded in Bayesian statistics. These intervals are more suitable for reliable post-data inference, particularly in KG accuracy evaluation. We prove that CrIs offer greater reliability and stronger guarantees than frequentist approaches in this context. Additionally, we introduce \emph{a}HPD, an adaptive algorithm that is more efficient for real-world KGs and statistically robust, addressing the interpretive challenges of CIs.
\end{abstract}



\received{17 October 2024}
\received[revised]{23 January 2025}
\received[accepted]{30 January 2025}

\maketitle

\acrodef{aHPD}[\textit{a}HPD]{\textit{adaptive} HPD}
\acrodef{AMT}[AMT]{Amazon Mechanical Turk}
\acrodef{AQP}[AQP]{Approximate Query Processing}
\acrodef{BET}[BET]{Belief Evaluation Task}
\acrodef{CEM}[CEM]{Cluster Error Model}
\acrodef{CDF}[CDF]{Cumulative Distribution Function}
\acrodef{CI}[CI]{Confidence Interval}
\acrodef{CLT}[CLT]{Central Limit Theorem}
\acrodef{CrI}[CrI]{Credible Interval}
\acrodef{CWA}[CWA]{Closed World Assumption}
\acrodef{DB}[DB]{Database}
\acrodef{DeFacto}[DeFacto]{Deep Fact Validation}
\acrodef{EI}[EI]{Entity Identification}
\acrodef{EL}[EL]{Entity Linking}
\acrodef{EPSEM}[EPSEM]{Equal Probability of Selection Method}
\acrodef{ER}[ER]{Entity Resolution}
\acrodef{ET}[ET]{Equal-Tailed}
\acrodef{FPC}[FPC]{Finite Population Correction}
\acrodef{FV}[FV]{Fact Verification}
\acrodef{GDA}[GDA]{Gene-Disease Association}
\acrodef{HPD}[HPD]{Highest Posterior Density}
\acrodef{ICC}[ICC]{Intracluster Correlation Coefficient}
\acrodef{IE}[IE]{Information Extraction}
\acrodef{IG}[IG]{Inference Graph}
\acrodef{iid}[i.i.d.]{independent and identically distributed}
\acrodef{IR}[IR]{Information Retrieval}
\acrodef{KGC}[KGC]{Knowledge Graph Construction}
\acrodef{KG}[KG]{Knowledge Graph}
\acrodef{LCWA}[LCWA]{Local Closed World Assumption}
\acrodef{LD}[LD]{Linked Data}
\acrodef{MC}[MC]{Monte Carlo}
\acrodef{MCTS}[MCTS]{Monte Carlo Tree Search}
\acrodef{MLSS}[MLSS]{Multi-Level Splitting Sampling}
\acrodef{MoE}[MoE]{Margin of Error}
\acrodef{NELL}[NELL]{Never Ending Language Learning}
\acrodef{OWA}[OWA]{Open World Assumption}
\acrodef{PDF}[PDF]{Probability Density Function}
\acrodef{QA}[QA]{Question Answering}
\acrodef{RCS}[RCS]{Random Cluster Sampling}
\acrodef{RE}[RE]{Relation Extraction}
\acrodef{RecSys}[RecSys]{Recommender Systems}
\acrodef{RS}[RS]{Reservoir Sampling}
\acrodef{SEM}[SEM]{Score Error Model}
\acrodef{SLSQP}[SLSQP]{Sequential Least SQuares Programming}
\acrodef{SRS}[SRS]{Simple Random Sampling}
\acrodef{SS}[SS]{Stratified Sampling}
\acrodef{STWCS}[STWCS]{Stratified TWCS}
\acrodef{TAA}[TAA]{Triples Accuracy Assessment}
\acrodef{TEM}[TEM]{Triple Error Model}
\acrodef{TWCS}[TWCS]{Two-stage Weighted Cluster Sampling}
\acrodef{WCS}[WCS]{Weighted Cluster Sampling}

\section{Introduction}
\label{sec:intro}

Over the past decade, the emergence of large-scale \acp{KG} like Wikidata~\cite{vrandecic_krotzsch-2014}, DBpedia~\cite{auer_etal-2007}, YAGO~\cite{hoffart_etal-2013}, and NELL~\cite{mitchell_etal-2018} has opened up new avenues in database research~\cite{ilyas_etal-2022} and empowered a range of machine learning applications, including virtual assistants~\cite{ilyas_etal-2023,mohoney_etal-2023}, search engines, recommender, and question answering systems~\cite{reinanda_etal-2020,samadi_etal-2015}. In this context, ensuring the accuracy of \ac{KG} content is pivotal to delivering a high-quality user experience, as emphasized by applications like Saga~\cite{ilyas_etal-2022}, where the on-demand evaluation of the \ac{KG} accuracy is a key feature.

The importance of assessing \ac{KG} accuracy is further accentuated by the limitations of current \ac{KG} construction processes, which often result in large volumes of incorrect facts, inconsistencies, and high sparsity~\cite{deshpande_etal-2013,pujara_etal-2017}. These issues not only undermine entity-oriented search applications by introducing potential misinformation~\cite{navalpakkam_etal-2013,hasibi_etal-2017,marchesin_etal-2024a}, but they also hinder the performance of downstream tasks such as \ac{KG} embeddings, which are integral to many machine learning applications~\cite{ilyas_etal-2023,mohoney_etal-2023}. \ac{KG} embeddings are sensitive to unreliable data and perform poorly on \acp{KG} extracted from noisy sources \cite{pujara_etal-2017}; thus, assessing the \ac{KG} accuracy is essential for making informed decisions about its use in both direct and downstream applications. 

Auditing \ac{KG} accuracy involves annotating facts with correctness labels. However, two major challenges arise when dealing with real-life \acp{KG}. First, obtaining high-quality labels is expensive~\cite{ojha_talukdar-2017,gao_etal-2019}. Secondly, annotating every fact in a large-scale \ac{KG} is unfeasible~\cite{qi_etal-2022}. Real-life \acp{KG} encompass hundreds of millions, or even billions, of relational facts, typically represented as $(s, p, o)$ triples. Therefore, cost-effective and scalable solutions are crucial. Given the importance of efficiently evaluating \ac{KG} accuracy, several studies have begun addressing these challenges~\cite{xue_zou-2023}, framing the \ac{KG} accuracy evaluation as a constrained minimization problem~\cite{gao_etal-2019, qi_etal-2022,marchesin_silvello-2024}. The state-of-the-art methods employ iterative procedures involving sampling strategies for efficient data collection, point estimators to determine accuracy, and $1-\alpha$ intervals to quantify the uncertainties associated with the sampling procedure. 

For $1-\alpha$ intervals, where $\alpha$ represents the risk of concluding that a difference exists between the estimated and true \ac{KG} accuracy, current approaches to \ac{KG} accuracy estimation rely on \acp{CI}~\cite{gao_etal-2019, qi_etal-2022,marchesin_silvello-2024}. \acp{CI} are rooted in frequentist statistics, where probability is interpreted as the long-term frequency of events over repeated sampling.
\rometa{Thus, the $1-\alpha$ probability pertains to the reliability of the interval estimation procedure (stochastic process) rather than to the specific interval computed given a sample (single event). Once a \ac{CI} is built from the sample, it either includes the true (but unknown) \ac{KG} accuracy or it does not -- it is no longer a matter of probability. However, in real-life scenarios, where \ac{KG} accuracy is typically assessed via a single, non-repeated process, there is instead a need for efficient $1-\alpha$ intervals that can be reliably interpreted in a \emph{one-shot} setting. To further reduce the appeal of \acp{CI} for the considered task, recent studies have also highlighted that their use can lead to interpretation fallacies regarding confidence, precision, and likelihood~\cite{morey_etal-2016}.} 

\paragraph{\textbf{Contributions}} \rometa{The core contributions of this work are:}
\begin{itemize}
    \item \rometa{We highlight the limitations of existing \ac{KG} evaluation methods relying on \acp{CI} and pioneer the use of Bayesian \acp{CrI} for \ac{KG} accuracy evaluation. By framing the task in Bayesian terms for the first time, we show that \acp{CrI} offer interpretable results at minimal costs, avoiding \ac{CI} interpretation fallacies and providing reliable solutions in a one-shot setting.}
    \item \rometa{We consider two main families of \acp{CrI}: \ac{ET} and \ac{HPD} intervals. We formally prove that \ac{HPD} \acp{CrI} provide the optimal representation of parameter values that are most consistent with the data across all annotation scenarios in \ac{KG} accuracy evaluation. This ensures comprehensive applicability across diverse \ac{KG} settings, using both informative and uninformative priors.} 
    \item To address the challenge of choosing the right prior, one of the most well-known resistances to using Bayesian methods in statistics, we introduce the \ac{aHPD} algorithm. \ac{aHPD} exploits multiple priors concurrently to generate different $1-\alpha$ \ac{HPD} intervals, competing to achieve convergence in the minimization problem. This ensures that the most efficient outcome is always obtained from competing solutions without choosing a specific prior ``a priori''.
    \item Finally, we show that \ac{aHPD} outperforms the state-of-the-art approaches with real and synthetic data at both small and large scales, proving robust regardless of the precision level required by the evaluation process and reducing annotation costs by up to $47\%$ in high-precision scenarios. 
\end{itemize}

\rometa{Although \ac{CI} fallacies and Bayesian \acp{CrI} are well-established concepts in statistical domains, they remain largely unfamiliar to the database community. This work bridges this gap by emphasizing their relevance for data quality management and providing a solution that addresses the interpretative fallacies currently affecting \ac{KG} quality estimation. Moreover, by removing the need for prior selection, our newly proposed \ac{aHPD} algorithm enables analysts to evaluate \ac{KG} accuracy with minimal statistical expertise. This makes the approach not only interpretable and efficient but also user-friendly. \ac{aHPD} is flexible and general enough to accommodate scenarios where (a) analysts have no prior knowledge about \ac{KG} accuracy, or (b) they possess reliable prior information, such as insights derived from similar \acp{KG}. This versatility ensures broad applicability while maintaining ease of use across diverse settings.} 

\paragraph{\textbf{Outline}} 
Section~\ref{sec:background} introduces the problem, its optimization, and the state-of-the-art sampling techniques and their corresponding estimators. Section~\ref{sec:confidence} discusses \acp{CI} and their limitations for \ac{KG} accuracy evaluation. Section~\ref{sec:credible} delves into \acp{CrI}, highlights the challenges of prior selection, and presents the \ac{aHPD} algorithm as an effective solution. Sections~\ref{sec:setup} and~\ref{sec:results} detail the experimental setup and results, respectively. Section~\ref{sec:relatedWork} reviews related work. Section~\ref{sec:conclusions} concludes the paper and outlines future research.

\section{Background}
\label{sec:background}

We present the problem, outline its optimization via an iterative evaluation framework, and report the state-of-the-art sampling techniques adopted in this work and the respective estimators.

\subsection{Preliminaries}
\label{subsec:prelim}

Following the notation introduced by~\citet{bonifati_etal-2018}, we consider \acp{KG} as ground RDF graphs defined as $G = (V, R, \eta)$, where $V = \{E \cup A\}$ is the set of nodes, $E$ the set of entities and $A$ of attributes. $R$ denotes the set of relationships between nodes and $\eta:R\rightarrow E \times \{E \cup A\}$ is the function that assigns an ordered pair of nodes to each relationship. The $\eta$ function generates the ternary relation $T$ of $G$, which consists of the set of $(s, p, o)$ triples where $s \in E$, $p \in R$, and $o \in \{E \cup A\}$, with a size $M = |T|$. Given $T$, we define an entity cluster $C_{e} = \{(s, p, o) \in T \ | \ s = e\}$ as the set of triples that share the same subject $e \in E$.

We treat triples as first-class citizens of a KG; hence, we redefine a \ac{KG} as $G = (V, R, T, \eta)$. A triple is also called a fact; we use the two terms interchangeably. 

\subsection{Problem: Constrained Minimization}
\label{subsec:problem}

Let the correctness of a triple $t \in T$ be represented by an indicator function $\mathds{1}(t)$, where $\mathds{1}(t) = 1$ indicates that the triple is correct, and $\mathds{1}(t) = 0$ indicates it is incorrect. The accuracy of a \ac{KG} can then be defined as the proportion of correct triples $\tau$ within $G$:
\begin{equation}
    \label{eq:kgAcc}
    \mu = \frac{\sum_{t \in T}\mathds{1}(t)}{M} = \frac{\tau}{M}
\end{equation}
where the value of $\mathds{1}(t)$ is determined through manual annotation. Following~\cite{gao_etal-2019,marchesin_silvello-2024}, we consider correctness a binary problem, similar to triple validation~\cite{esteves_etal-2018}, since an atomic fact is correct or incorrect.

However, manually evaluating every triple in a large-scale \ac{KG} to determine its accuracy is impractical. Therefore, the standard approach is to estimate $\mu$ using an estimator $\hat{\mu}$, calculated from a relatively small sample drawn according to a sampling strategy $\mathcal{S}$, which selects a subset $T_{\mathcal{S}} \subset T$ of triples to annotate. This results in a sample $G_{\mathcal{S}} = (V_{\mathcal{S}}, R_{\mathcal{S}}, T_{\mathcal{S}}, \eta)$, where $V_{\mathcal{S}} \subset V$ and $R_{\mathcal{S}} \subset R$.

To ensure the accuracy of $G$, the estimator $\hat{\mu}$ must be unbiased, meaning $E[\hat{\mu}] = \mu$. Additionally, since $\hat{\mu}$ is a point estimator, it is necessary to provide a $1-\alpha$ interval to quantify the uncertainties associated with the sampling procedure. The larger the sample $G_{\mathcal{S}}$, the narrower the interval until it reaches zero width when the sample is equivalent to $G$. A key measure related to intervals is the \ac{MoE}, which is half the width of the interval.

Let $G_{\mathcal{S}} = \mathcal{S}(G)$ be a sample drawn using the sampling strategy $\mathcal{S}$, and $\hat{\mu}$ an estimator of $\mu$ based on that sample. Let $\operatorname{cost}(G_{\mathcal{S}})$ denote the cost of manually evaluating the correctness of elements in the sample. Accordingly to~\cite{gao_etal-2019,marchesin_silvello-2024}, we define the \ac{KG} accuracy evaluation task as a constrained minimization problem:

\begin{problem}
\label{def:problem}
Given a \ac{KG} $G$ and an upper bound $\varepsilon$ for the \ac{MoE} of a $1-\alpha$ interval:
\begin{equation*}
\begin{aligned}
\underset{\mathcal{S}}{\operatorname{minimize}} \quad & \operatorname{cost}(G_{\mathcal{S}}) \\
\operatorname{subject \ to} \quad & E[\hat{\mu}] = \mu, \text{MoE} \leq \varepsilon
\end{aligned}
\end{equation*}
\end{problem}

\subsection{Optimization: Iterative Procedure}
\label{subsec:optimization}

\begin{figure}[t!]
\centering
\includegraphics[width=\linewidth]{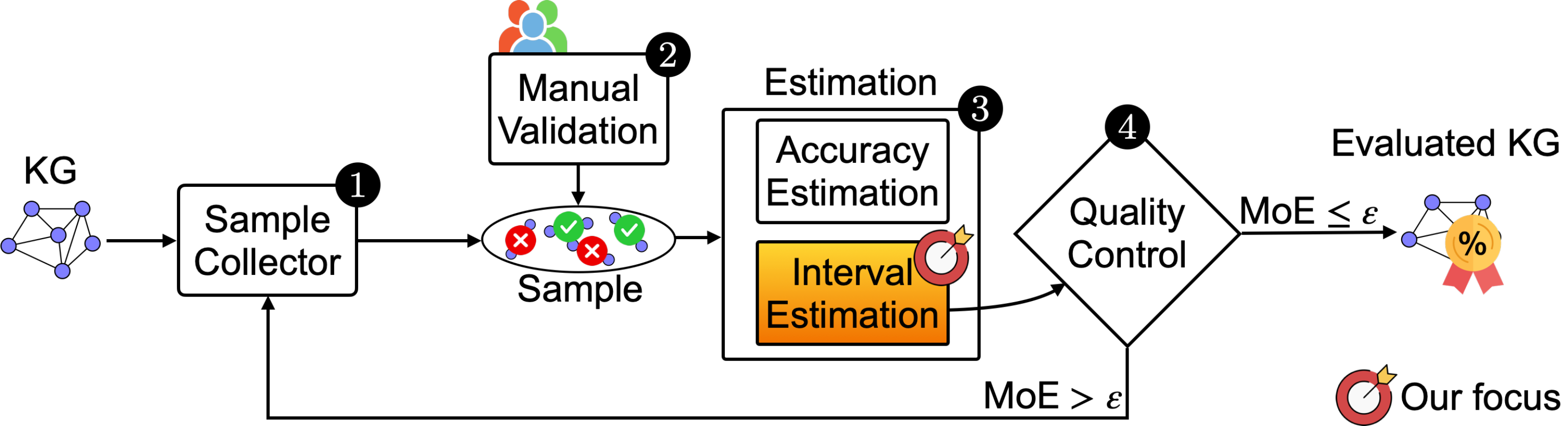}
  \caption{Efficient KG accuracy evaluation framework~\cite{marchesin_silvello-2024}.}
  \label{fig:framework}
\end{figure}

The minimization problem is optimized with an evaluation framework that operates as the iterative procedure shown in Figure~\ref{fig:framework}. 

In phase \circled{1}, a small batch of triples is drawn from the \ac{KG} according to a chosen sampling strategy $\mathcal{S}$. This strategy should be selected to minimize the cost function, representing the optimization objective. In phase \circled{2}, manual annotations are obtained for these newly sampled triples and combined with any existing annotations. In phase \circled{3}, using the annotations accumulated via the chosen sampling design, an unbiased estimator computes $\hat{\mu}$, the estimate of the \ac{KG} accuracy. The annotated sample also constructs the corresponding $1-\alpha$ interval, quantifying the uncertainties in the sampling procedure. In phase \circled{4}, a quality control step checks whether the generated interval meets the predefined upper bound, ensuring that $\text{MoE} \leq \varepsilon$. If the criterion is satisfied, the process concludes, and the accuracy estimate and the corresponding $1-\alpha$ interval are reported. If not, the process returns to phase \circled{1}. Thus, the framework iteratively samples and estimates, stopping when the interval meets the specified threshold $\varepsilon$. Doing so avoids oversampling and unnecessary manual annotations, ensuring accurate and cost-effective estimates.

The fact that the problem is solved once the interval is sufficiently narrow highlights the pivotal role of $\mathbf{1-\alpha}$ \textbf{intervals} in the optimization process. Given a sampling strategy, faster-shrinking intervals allow the problem to be solved with smaller samples, thereby reducing the number of required annotations compared to other, slower-to-converge intervals. At the same time, it is also essential that $1-\alpha$ intervals are reliable and guarantees the statistical properties they (nominally) entail, as first outlined in~\cite{marchesin_silvello-2024}. However, to ensure reliability in the evaluation process, the approach by \citet{marchesin_silvello-2024} required a trade-off, thus sacrificing some efficiency. In this work, by shifting the interval estimation framework from frequentist to Bayesian statistics, we demonstrate that it is possible to obtain intervals that are guaranteed to be the shortest -- and thus the most efficient -- while also providing probabilistic solid guarantees on their reliability.

\subsection{Sampling and Accuracy Estimation}
\label{subsec:sampling}

The approaches proposed to efficiently evaluate \ac{KG} accuracy~\cite{gao_etal-2019,qi_etal-2022,marchesin_silvello-2024} adopt simple and cluster random sampling techniques, paired with well-known point estimators~\cite{cochran-1977}. Below, we detail both sampling strategies and the corresponding unbiased estimators.

\paragraph{\textbf{Simple Random Sampling}}

\ac{SRS} draws a sample of $n_{\mathcal{S}}$ triples from $G$ without replacement. In large-scale \acp{KG}, the probability of choosing the same triple more than once is low, making sampling with replacement a good approximation to sampling without replacement~\cite{casella_berger-2002} and a practical solution. 

Under \ac{SRS}, the estimator for $\mu$ is the sample proportion:
\begin{equation}
    \label{eq:srs}
    \hat{\mu}_{\text{\scriptsize SRS}} = \frac{\sum_{i=1}^{n_{\mathcal{S}}} \mathds{1}(t_{i})}{n_{\mathcal{S}}} = \frac{\tau_{\mathcal{S}}}{n_{\mathcal{S}}}
\end{equation}
with estimation variance given by $V(\hat{\mu}_{\text{\scriptsize SRS}})= \frac{\hat{\mu}_{\text{\scriptsize SRS}}(1-\hat{\mu}_{\text{\scriptsize SRS}})}{n_{\mathcal{S}}}$.

When \ac{SRS} is applied, $\hat{\mu}_{\text{\scriptsize SRS}}$ is an unbiased estimator of $\mu$~\cite{cochran-1977}. 

\paragraph{\textbf{Cluster Random Sampling}}

Cluster sampling is a cost-effective alternative to \ac{SRS} when estimating the accuracy of large-scale \acp{KG}~\cite{gao_etal-2019,qi_etal-2022,marchesin_silvello-2024}. Among the different cluster sampling strategies that can be employed, \ac{TWCS} represents the state-of-the-art for \ac{KG} accuracy evaluation~\cite{gao_etal-2019,marchesin_silvello-2024}. \ac{TWCS} is divided into two stages. In the first stage, \ac{TWCS} samples $n_{C}$ entity clusters with probabilities $\pi_{i}$ proportional to their sizes. That is, by denoting the size of the $i$th entity cluster as $M_{i} = |C_{e_{i}}|$, the cluster selection probability can be defined as $\pi_{i} = M_{i}/M$. In the second stage, \ac{TWCS} draws $\min\{M_{i}, m\}$ triples from each sampled cluster using \ac{SRS} without replacement.

Let $\hat{\mu}_{i}$ be the (estimated) accuracy of the $i$th entity cluster. The estimator for $\mu$ under \ac{TWCS} is:
\begin{equation}
    \label{eq:twcs}
    \hat{\mu}_{\text{\scriptsize TWCS}} = \frac{\sum_{i=1}^{n_{C}}\hat{\mu}_{i}}{n_{C}}
\end{equation}
with estimation variance $V(\hat{\mu}_{\text{\scriptsize TWCS}}) = \frac{1}{n_{C}(n_{C}-1)}\sum_{i=1}^{n_{C}}(\hat{\mu}_{i} - \hat{\mu}_{\text{\scriptsize TWCS}})^{2}$.

When \ac{TWCS} is applied with a second-stage sample of approximately size $m$, $\hat{\mu}_{\text{\scriptsize TWCS}}$ is an unbiased estimator of $\mu$~\cite{cochran-1977}.

\section{State of the Art: Confidence Intervals (CIs)}
\label{sec:confidence}

Previous research on efficient \ac{KG} accuracy evaluation has utilized \acp{CI} to estimate $1-\alpha$ intervals~\cite{gao_etal-2019,qi_etal-2022,marchesin_silvello-2024}. However, no universal formula exists for constructing \acp{CI} under arbitrary sampling strategies (and point estimators). Instead, multiple approaches have been developed~\cite{brown_etal-2001}, each offering \acp{CI} with distinct properties~\cite{morey_etal-2016}. Among these, we review only the methods employed by state-of-the-art \ac{KG} accuracy estimation techniques — specifically, the Wald and Wilson approaches~\cite{casella_berger-2002,wilson-1927}.

This section outlines Wald and Wilson methods for constructing $1-\alpha$ \acp{CI}. Next, we discuss the common misinterpretations and the resulting fallacies regarding \acp{CI} concerning their confidence, precision, and likelihood properties~\cite{morey_etal-2016}, which are desirable characteristics for tasks where analysts seek reasonable post-data inference, such as in KG accuracy estimation.

\subsection{The Wald Method}
\label{subsec:wald}

The Wald method relies on normal approximation to build $1-\alpha$ \acp{CI} and involves inverting the acceptance region of the Wald large-sample normal test~\cite{casella_berger-2002}: 
\begin{equation}
    \label{eq:waldTest}
    \left|\frac{\hat{\mu}_{\mathcal{S}}-\mu}{\sqrt{V(\hat{\mu}_{\mathcal{S}})}}\right| \leq z_{\alpha/2}
\end{equation}
where $\mu$ represents the true \ac{KG} accuracy, $\hat{\mu}_{\mathcal{S}}$ and $V(\hat{\mu}_{\mathcal{S}})$ the estimated \ac{KG} accuracy and its variance under a given sampling strategy $\mathcal{S}$, and $z_{\alpha/2}$ the critical value of the standard normal distribution for a confidence level $1-\alpha$.

By solving Equation~(\ref{eq:waldTest}) in $\mu$, we obtain the $1-\alpha$ Wald interval: 
\begin{equation}
    \label{eq:waldInt}
    \hat{\mu}_{\mathcal{S}} \pm z_{\alpha/2} \sqrt{V(\hat{\mu}_{\mathcal{S}})}
\end{equation}

The Wald interval is an appealing \ac{CI} due to its simplicity. However, when applied to binomial proportions, it is known to incur overshooting and zero-width intervals~\cite{brown_etal-2001}, also presenting an erratic behavior when used to gauge \ac{KG} accuracy \cite{marchesin_silvello-2024}. These phenomena aggravate in \acp{KG} with accuracy values deviating from the center of the (accuracy) range -- where normal approximation works best -- towards its boundaries, that is, one or zero. Since many real-life \acp{KG}, such as YAGO, DBpedia, and Wikidata, involve a certain degree of manual curation, they rarely have accuracy values as low as $0.5$. Conversely, fully automated \acp{KG}, which are noisier and more error-prone, frequently present accuracy values well below $0.5$~\cite{pujara_etal-2017}. Thus, this tendency of real-life \acp{KG} towards the boundaries of the accuracy range makes Wald an often unreliable choice for \ac{KG} accuracy estimation. 

\subsection{The Wilson Method}
\label{subsec:wilson}

To overcome Wald limitations, \citet{marchesin_silvello-2024} relied on the Wilson method~\cite{wilson-1927}, which can be used to build $1-\alpha$ binomial \acp{CI}. Similarly to Wald, the method involves inverting the acceptance region of the significance test in Equation~\ref{eq:waldTest}, but requires two changes. First, it requires assuming that the sampling strategy used to collect the sample is \ac{SRS}. Secondly, it requires replacing the estimated standard error with the null standard error. Based on these changes, the test can thus be rewritten as:
\begin{equation}
    \label{eq:wilsonTest}
    \left|\sqrt{\frac{n_{\mathcal{S}}}{\mu(1-\mu)}}\cdot (\hat{\mu}_{\text{\scriptsize SRS}}-\mu)\right| \leq z_{\alpha/2}
\end{equation}

By solving Equation~(\ref{eq:wilsonTest}) in $\mu$, we get the $1-\alpha$ Wilson interval:
\begin{equation}
    \label{eq:wilsonInt}
    \frac{\hat{\mu}_{\text{\scriptsize SRS}} + \frac{z_{\alpha/2}^2}{2n_{\mathcal{S}}}}{1 + \frac{z_{\alpha/2}^2}{n_{\mathcal{S}}}} \pm \frac{z_{\alpha/2}}{1 + \frac{z_{\alpha/2}^2}{n_{\mathcal{S}}}} \cdot \sqrt{\frac{\hat{\mu}_{\text{\scriptsize SRS}}(1-\hat{\mu}_{\text{\scriptsize SRS}})}{n_{\mathcal{S}}} + \frac{z_{\alpha/2}^2}{4n_{\mathcal{S}}^{2}}}
\end{equation}
which, compared to the Wald interval, consists of a relocated center estimate (left) and a corrected standard deviation (right). 

Wilson intervals correct the erratic behavior of Wald ones, resulting in higher reliability, although at the expense of a drop in efficiency as accuracy approaches its boundaries~\cite{marchesin_silvello-2024}. Note that design effect adjustments are also required to adapt Wilson to complex sampling strategies~\cite{kish-1965,kish-1995}. 

\subsection{Confidence Intervals (CIs) Limitations}
\label{subsec:ciLimitations}

\acp{CI} fall under frequentist statistics, where probability is interpreted as the long-term frequency of events across repeated sampling. This means that, when building a \ac{CI}, we are not directly estimating the probability that the parameter of interest -- the \ac{KG} accuracy in our case -- lies within the interval. Instead, we estimate how frequently such intervals, across repeated trials, will catch the true parameter. This is often referred to as \emph{frequency probability}.

Hence, \acp{CI} interpretation can be counterintuitive and, for this reason, is often misunderstood by practitioners~\cite{hoekstra_etal-2014}. Three major fallacies exist in interpreting \acp{CI}, as shown by ~\citet{morey_etal-2016}. \rometa{Although these fallacies do not affect the quantitative outcomes of \ac{KG} accuracy evaluation, they pose a significant risk to the proper use of \acp{CI}. Misinterpretations stemming from them can lead to conclusions as flawed as those caused by quantitative inaccuracies.} 
\rometa{In the following, we first outline these fallacies and then provide a practical example where they occur.}

\begin{fallacy}
    \label{fall:1}
    The probability that a $1-\alpha$ \ac{CI} contains the true parameter is $1-\alpha$ for a particular observed interval.
\end{fallacy}

This is a fallacious interpretation because once a CI is constructed, the probability that it contains the true parameter is either 1 (if it does) or 0 (if it does not). The $1-\alpha$ confidence level represents the reliability of the interval estimation procedure across repeated samples rather than the probability that any specific interval captures the true parameter. This misconception stems from the belief that a $1-\alpha$ CI provides the likelihood that the unknown parameter falls within the observed interval~\cite{hoekstra_etal-2014}. However, the $1-\alpha$ confidence level does not assess whether the specific interval computed from the data includes the true parameter. CIs rely on pre-data assumptions not intended for post-data inference~\cite{neyman_harold-1937}.

\begin{fallacy}
    \label{fall:2}
    The width of a \ac{CI} reflects the precision of our knowledge about the parameter.
\end{fallacy}

This interpretation is flawed because there is no inherent relationship between the width of a CI and the precision of the parameter estimate. The $1-\alpha$ confidence level reflects the reliability of the interval estimation procedure, not the precision of a particular observed interval. Hence, it is impossible to know whether a specific narrow or wide interval belongs to the $1-\alpha$ proportion that reliably captures the true parameter or the $\alpha$ proportion that does not. Consequently, interpreting CI width as a measure of precision can lead to incorrect post-data conclusions~\cite{morey_etal-2016}.

\begin{fallacy}
    \label{fall:3}
    A \ac{CI} includes the likely values of the parameter. 
\end{fallacy}

While a CI is constructed to have a fixed $1-\alpha$ probability of capturing the true parameter across repeated samples, this does not imply that any particular observed interval contains the most likely values for the parameter. Once the interval is calculated, it either includes the true parameter or does not~\cite{neyman-1941}. Indeed, CIs can sometimes exclude many reasonable values or, in extreme cases, be infinitesimally narrow or even empty (e.g., Wald intervals~\cite{brown_etal-2001}), thereby excluding all possible values~\cite{morey_etal-2016}.

\begin{example}
\label{ex:ciFallacies}
\rometa{Let us consider an analyst who wants to estimate the accuracy of NELL ($\mu = 0.91$, see Table~\ref{tab:kgStats}) using the evaluation framework described in Section~\ref{subsec:optimization}. The analyst adopts the most straightforward approach with \ac{SRS} for sampling, the Wald method with a significance level $\alpha=0.05$ for interval estimation, and a \ac{MoE} threshold of $\varepsilon = 0.05$ for convergence. Under this setup, the evaluation procedure halts (i.e., \ac{MoE} $\leq \varepsilon$) with a sample size of $n_\mathcal{S} = 30$ and an estimated accuracy of $\hat{\mu}_{\text{\scriptsize SRS}} = 1.00$ -- reasonable outcomes given the underlying \ac{KG} accuracy.\footnote{\rometa{In our experiments, these results occurred in $7\%$ of the $1,000$ iterations run on NELL.}} With these values, the estimation variance is $V(\hat{\mu}_{\text{\scriptsize SRS}}) = 0.00$. Substituting this into Equation~(\ref{eq:waldInt}) produces a zero-width interval: $\text{CI} = [1.00, 1.00]$, meaning that there is an absolute certainty that the estimated accuracy $\hat{\mu}_{\text{\scriptsize SRS}}$ is correct. However, this also implies that the probability that the obtained \ac{CI} contains the true accuracy cannot be $1-\alpha$. Therefore, placing $1-\alpha$ confidence in an interval that assumes absolute certainty would clearly be a mistake for the analyst (\textbf{Fallacy 1}). At the same time, a zero-width \ac{CI} does not imply that effect size is determined with perfect precision, as that would require annotating the entire population, nor can it imply that there is a $0.95$ ($1-\alpha$) probability that $\mu$ is exactly $1.00$, thus leading to \textbf{Fallacy 2}. Finally, a zero-width interval excludes all the possible values the accuracy can take, resulting in \textbf{Fallacy 3}. Hence, these fallacies highlight that the post-data properties required to make valid inferences from the interval are absent, potentially leading to arbitrary conclusions.}
\end{example}

\paragraph{\textbf{Task-Specific Deterrents}}

The frequentist assumption underlying \acp{CI} -- and the corresponding interpretation fallacies this can generate -- makes these intervals not ideal for \ac{KG} accuracy evaluation. In fact, \acp{CI} offer guarantees over repeated trials rather than for a specific instance. However, in real-life scenarios, \ac{KG} accuracy evaluation is typically conducted as a single, non-repeated process. Hence, \acp{CI} are inconvenient and $1-\alpha$ intervals providing a reliable interpretation in a one-shot setting are very much desired.

Moreover, the lack of a unified framework for constructing CIs contributes to their complexity. Different CIs are generated using distinct construction methods, often relying on intricate procedures that involve assumptions and approximations that are rarely tested~\cite{makowski_etal-2019}. Additionally, CIs constructed through different methods can vary in efficiency and reliability across different regions of the accuracy space~\cite{marchesin_silvello-2024}. Therefore, establishing a unified framework that enables the seamless computation of multiple $1-\alpha$ intervals would be crucial for improving consistency and comparability.

Finally, the long-term properties of CIs necessitate additional metrics to assess their reliability, such as coverage probability~\cite{marchesin_silvello-2024}. However, calculating this probability requires repeated iterations of the entire KG accuracy evaluation procedure, making it impractical for real-world applications. 

\section{Proposed Solution: Credible Intervals (CrIs)}
\label{sec:credible}

To overcome the limitations of \acp{CI} for \ac{KG} accuracy evaluation, we can resort to \acfp{CrI}~\cite{edwards_etal-1963}. These intervals are grounded in Bayesian statistics and operate on the principle that estimated parameters are random variables following a distribution. 

Bayesian statistics differs from the frequentist approach as it is a post-data theory~\cite{morey_etal-2016}. It uses the information from the data, combined with model assumptions and prior information, to determine what is reasonable to believe~\cite{gelman-2008,wasserman-2008}.

In the Bayesian framework, probability represents a degree of belief in an event, captured by a distribution. Before observing data, this belief is expressed through a prior distribution, reflecting initial assumptions or knowledge about the parameter of interest. For instance, when estimating \ac{KG} accuracy, the prior distribution encodes beliefs about likely accuracy values based on prior knowledge or judgment. Once data is collected, this initial belief is updated through the likelihood, representing the posterior probability -- the focus of Bayesian statistics -- from the observed data. This probability is obtained by conditioning the prior probability with the information summarized by the likelihood via Bayes' rule~\cite{box_tiao-2011}. 

A \ac{CrI} is an interval derived from the posterior distribution, indicating the range within which an unobserved parameter value will likely fall with a certain probability. Since \acp{CrI} are defined on the posterior distribution, a $1-\alpha$ \ac{CrI} has the following properties: 
\begin{enumerate}
\item It contains the parameter of interest with $1-\alpha$ probability (addressing Fallacy~\ref{fall:1}).
\item Its width reflects the precision of the estimate (addressing Fallacy~\ref{fall:2}).
\item It covers the plausible values of the parameter (addressing Fallacy~\ref{fall:3}). Thus, Bayesian interval estimation avoids CIs interpretation fallacies by providing $1-\alpha$, interpretable and reliable intervals in a one-shot setting.
\end{enumerate}

We introduce two families of \acp{CrI}: \ac{ET} and \ac{HPD} intervals~\cite{box_tiao-2011}. We demonstrate that \ac{HPD} intervals are the shortest possible, providing the best representation of parameter values most consistent with the data. Additionally, we show that \ac{ET} and \ac{HPD} intervals are equivalent when the posterior distribution is symmetric. Finally, we address the challenges of selecting an appropriate prior by proposing a new adaptive solution that eliminates the need for manual prior selection while ensuring optimal performance.

\subsection{Bayesian Interval Estimation}
\label{subsec:bayesFramework}

We model the process of annotating triples for correctness as a binomial process $\operatorname{Bin}(n_{\mathcal{S}}, \mu)$, where $n_{\mathcal{S}} \geq 1$ is the number of annotated triples and $\mu$ represents the binomial proportion (cf. Eq.~(\ref{eq:kgAcc})). Since beta distributions are the standard conjugate priors for binomial distributions, using them for inference on $\mu$ is suitable~\cite{berger-1985}. 

Let $\tau_{\mathcal{S}} \sim \operatorname{Bin}(n_{\mathcal{S}}, \mu)$, and assume that $\mu$ has a prior distribution $\operatorname{Beta}(a, b)$, where $a, b > 0$ represent our prior belief regarding correct ($a$) and incorrect ($b$) triples in the \ac{KG}. Due to the conjugate relationship between beta and binomial distributions, the posterior distribution of $\mu$ can be derived as $\operatorname{Beta}(a+\tau_{\mathcal{S}},b+n_{\mathcal{S}}-\tau_{\mathcal{S}})$~\cite{raiffa_schlaifer-1961}. This means that to obtain the beta posterior, we simply update the parameters of the beta prior by adding the observed number of correct ($\tau_{\mathcal{S}}$) and incorrect ($n_{\mathcal{S}}-\tau_{\mathcal{S}}$) triples.

Let $F(\mu~|~G_{\mathcal{S}})$ denote the posterior \ac{CDF} of $\mu$. For a given $\alpha$, we can identify an interval $(l, u)$ such that:
\begin{equation}
    \label{eq:crInt}
        1-\alpha = \operatorname{Pr}(l  \leq  \mu  \leq  u~|~G_{\mathcal{S}}) = F(u) - F(l)
\end{equation}
Therefore, a $1-\alpha$ \ac{CrI} is an interval $(l, u)$ that satisfies the condition $1-\alpha = F(u) - F(l)$~\cite{press-2002}. However,it remains to decide how to compute such an interval.

\subsection{Equal-Tailed (ET) Intervals}
\label{subsec:etInt}

A common solution is that of \acf{ET} intervals~\cite{mcelreath-2020}. These are $1-\alpha$ \acp{CrI} that allocate equal probability to the tails of the posterior distribution -- i.e., below the lower bound $l$ and above the upper bound $u$. In other words, a $1-\alpha$ \ac{ET} \ac{CrI} is obtained by taking the central $1-\alpha$ region of the posterior distribution, leaving $\alpha/2$ probability in each tail. By denoting $\operatorname{qBeta}(i; \cdot, \cdot)$ as the $i$th quantile of a beta distribution, we define the $1-\alpha$ \ac{ET} \ac{CrI} as:
\begin{equation}
    \label{eq:etInt}
    \begin{split}
    & l = \operatorname{qBeta}(\alpha/2; a + \tau_{\mathcal{S}}, b + n_{\mathcal{S}} - \tau_{\mathcal{S}}) \\
    & u = \operatorname{qBeta}(1-\alpha/2; a + \tau_{\mathcal{S}}, b + n_{\mathcal{S}} - \tau_{\mathcal{S}})
    \end{split}
\end{equation}

\begin{figure}[t!]

\centering
\includegraphics[width=\linewidth]{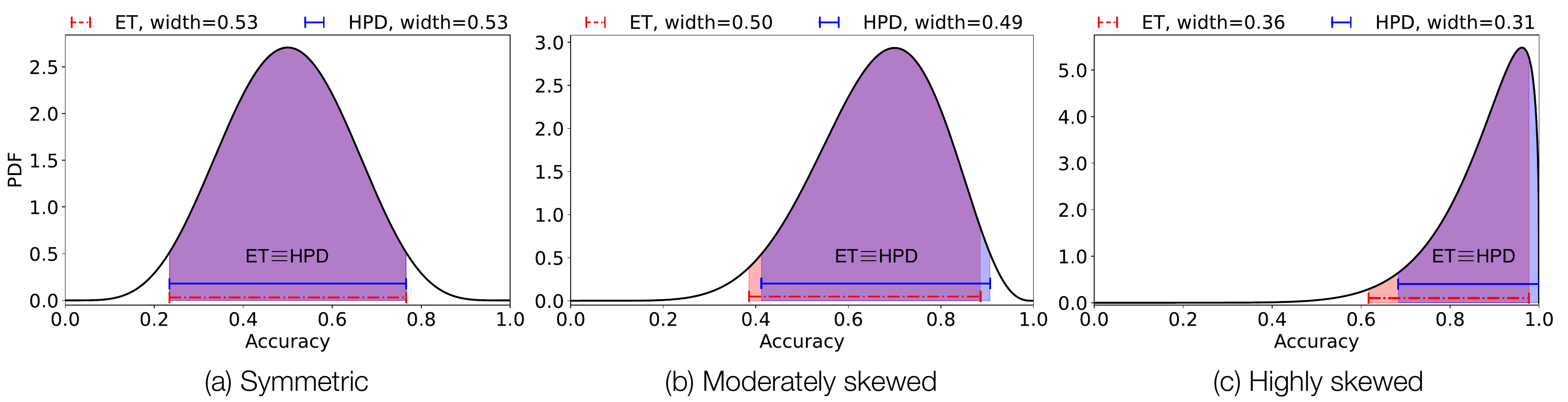}
  \caption{\rometa{Comparison of \ac{ET} and \ac{HPD} \acp{CrI} across three posterior distributions for \ac{KG} accuracy with increasing skewness (from left to right). In panel (a), where the posterior is symmetric, both \ac{ET} and \ac{HPD} intervals capture the most probable values (purple region). However, as skewness increases in panels (b) and (c), the \ac{ET} interval deviates from the highest posterior density region, including less likely parameter values (red region), and resulting in unnecessarily longer intervals to satisfy the $1-\alpha = F(u) - F(l)$ condition. In contrast, the \ac{HPD} intervals remain optimal, covering only the highest posterior density region (purple and blue regions), thereby highlighting the superior performance of \ac{HPD} in skewed distributions compared to \ac{ET}.}}
  \label{fig:etVShpd}

\end{figure}

While \ac{ET} \acp{CrI} are intuitive and effective for symmetric posteriors, they become suboptimal for skewed distributions because, by design, they allocate $\alpha/2$ of the distribution in each tail. To illustrate this, consider Figure~\ref{fig:etVShpd}, which shows three scenarios with increasing skewness in the posterior distribution for \ac{KG} accuracy. In the symmetric case (Figure~\ref{fig:etVShpd}(a)), the \ac{ET} \ac{CrI} captures the most probable values \rometa{(purple region)}, providing a good summary of the distribution. However, as skewness increases (Figures~\ref{fig:etVShpd}(b,c)), the \ac{ET} \ac{CrI} includes lower-probability parameter values \rometa{(red region) while excluding parts of the \ac{HPD} region (blue region)}. This results in a longer interval than necessary to satisfy the condition $1-\alpha = F(u) - F(l)$. 

\rometa{To further support this, we compared the \ac{CDF} of the \ac{HPD} region excluded by the \ac{ET} \ac{CrI} with the \ac{CDF} of any equally wide region covered by \ac{ET} but falling outside the \ac{HPD} region.  In the moderately skewed case (Figure ~\ref{fig:etVShpd}(b)), the CDF of these regions falling outside the HPD region, but covered by the ET CrI, is always less than 75\% of the CDF of the HPD region not covered by the ET CrI. In the highly skewed case (Figure~\ref{fig:etVShpd}(c)), the \ac{CDF} of these non-\ac{HPD} regions is not even the $20\%$ of the CDF of the excluded \ac{HPD} region. We can see that, in both cases, the exclusion of portions of the \ac{HPD} region forces the \ac{ET} interval to expand in width to satisfy the $1-\alpha = F(u) - F(l)$ constraint, making it a suboptimal choice.}

\rometa{Moreover, for skewed posteriors, the inability of \ac{ET} \acp{CrI} to fully cover the \ac{HPD} region makes them vulnerable to Fallacy~\ref{fall:3}.} Thus, for \ac{KG} accuracy evaluation, where skewed distributions are common, \ac{ET} limitations are of significance.

\subsection{Highest Posterior Density (HPD) Intervals}
\label{subsec:hpdInt}

The $1-\alpha$ \ac{CrI} that best describes the posterior distribution is the interval containing the most probable posterior values~\cite{mcelreath-2020}, i.e., the \acf{HPD} interval. 

\paragraph{\textbf{Definition.}} Let $f(\mu|G_{\mathcal{S}})$ represent the \ac{PDF} of the posterior distribution. A $1-\alpha$ \ac{CrI} is an \ac{HPD} interval if it satisfies the following two conditions:
\begin{enumerate}
        \item[(1)] $F(u) - F(l) = 1 - \alpha$;
        \item[(2)] For $l  \leq  \mu  \leq  u$, $f(\mu|G_{\mathcal{S}})$ is greater than for any other interval where condition (1) holds.
\end{enumerate}
The first condition corresponds to the general definition of a $1-\alpha$ \ac{CrI} (see Eq.~(\ref{eq:crInt})), while the second ensures that every point inside the \ac{HPD} interval has higher \ac{PDF} value than points outside.
Given these conditions, within the Bayesian framework designed in Section~\ref{subsec:bayesFramework}, the \ac{HPD} interval is the smallest possible and unique. \rometa{As we will see in Section~\ref{sec:results}, these properties translate into more cost-effective solutions for \ac{KG} accuracy evaluation.}

First, we prove both properties for the \textbf{standard case},  where the number of correct ($\tau_\mathcal{S}$) and incorrect ($n_{\mathcal{S}} - \tau_{\mathcal{S}}$) triples resulting from the annotation process are nonzero. \rtwo{Due to space limits, we report sketch proofs. The complete proofs are in the online repository.\footnote{\rtwo{\url{https://github.com/KGAccuracyEval/credible-intervals4kg-estimation}}}}
 
\begin{theorem}
    \label{thm:hpdMin}
    Given a prior distribution $\operatorname{Beta}(a, b)$ for $\mu$ and a \ac{KG} correctness annotation process, where $0 < \tau_{\mathcal{S}} < n_{\mathcal{S}}$, the $1-\alpha$ \ac{HPD} interval is the smallest $(l, u)$ interval satisfying the condition $F(u) - F(l) = 1 - \alpha$.
\end{theorem}

\begin{proof}[\rtwo{Sketch Proof}]
    \rtwo{When $0 < \tau_{\mathcal{S}} < n_{\mathcal{S}}$, the beta posterior $\operatorname{Beta}(a+\tau_{\mathcal{S}},b+n_{\mathcal{S}}-\tau_{\mathcal{S}})$ is unimodal and continuous over $[0, 1]$, ensuring the existence of a \ac{CrI}. To find the smallest interval satisfying $F(u) - F(l) = 1 - \alpha$, we apply the method of Lagrange multipliers, which minimizes the interval length $(u-l)$ under the constraint that the posterior probability over $[l, u]$ equals $1-\alpha$. That is, we minimize the Lagrangian function $\mathcal{L} = (u - l) + \lambda\Bigl(\int_{l}^{u} f(\mu~|~G_{\mathcal{S}})~d\mu - (1-\alpha)\Bigr)$, where $\lambda$ is the Lagrange multiplier.}

    \rtwo{From the first-order conditions $\partial \mathcal{L}/\partial l = 0$ and $\partial \mathcal{L}/ \partial u = 0$, we derive $f(l~|~G_{\mathcal{S}}) = f(u~|~G_{\mathcal{S}}) = -\frac{1}{\lambda}$. We compute the second-order derivatives to verify that $(l, u)$ minimizes $\mathcal{L}$. The unimodal nature of the posterior ensures that the derived Hessian matrix is positive definite, confirming that $(l, u)$ is the smallest interval satisfying $F(u) - F(l) = 1 - \alpha$.}
\end{proof}

\begin{theorem} 
    \label{thm:hpdUnique}
    Under the assumptions of Theorem~\ref{thm:hpdMin}, the $1-\alpha$ \ac{HPD} interval is unique. 
\end{theorem}

\begin{proof}[\rtwo{Sketch Proof}]
    \rtwo{Let $(l^*, u^*)$ represent the \ac{HPD} interval. Assume by contradiction that there exists another interval $(l', u') \neq (l^*, u^*)$ such that $(u'-l') = (u^* - l^*)$ and $\int_{l'}^{u'} f(\mu~|~G_{\mathcal{S}})~d\mu = 1 - \alpha$. Since the posterior is unimodal, Th.~\ref{thm:hpdMin} ensures that the \ac{HPD} interval contains the mode. Similarly, the alternative interval $(l', u')$ must also contain the mode. Hence, given that $(l', u') \neq (l^{*}, u^{*})$ but $(u' - l') = (u^{*} - l^{*})$, two cases arise: (i) $l' < l^{*}$ and $u' < u^{*}$; or (ii) $l' > l^{*}$ and $u' > u^{*}$. We prove case (i); case (ii) is analogous.}

    \rtwo{Since both intervals satisfy the condition $F(u) - F(l) = 1-\alpha$, then it must hold that $\int_{l'}^{u'}f(\mu~|~G_{\mathcal{S}})~d\mu$ $= \int_{l^{*}}^{u^{*}}f(\mu~|~G_{\mathcal{S}})~d\mu$. Removing the common terms yields: $\int_{l'}^{l^{*}}f(\mu~|~G_{\mathcal{S}})~d\mu = \int_{u'}^{u^{*}}f(\mu~|~G_{\mathcal{S}})~d\mu$. However, since $l'$ lies outside the \ac{HPD} region while $u'$ lies inside, it follows that $\int_{l'}^{l^{*}}f(\mu~|~G_{\mathcal{S}})~d\mu < \int_{u'}^{u^{*}}f(\mu~|~G_{\mathcal{S}})~d\mu$, which contradicts the equality assumption.}
\end{proof}

To compute the $1-\alpha$ \ac{HPD} \ac{CrI}, we need to find the $(l, u)$ interval that minimizes the Lagrangian function $\mathcal{L}$ defined in Th.~\ref{thm:hpdMin}. We resort to sequential quadratic programming by adopting the \ac{SLSQP} optimizer~\cite{kraft-1988}, that minimizes a Lagrangian of several variables with any combination of bounds, equality, and inequality constraints. \ac{SLSQP} reformulates the optimization problem into a sequence of quadratic subproblems whose objectives are second-order approximations of the original Lagrangian and whose constraints are linearized versions of the original ones. The method then applies globalization techniques to guarantee convergence.

In our case, the objective of the Lagrangian is minimizing the $(l, u)$ interval width, with $l$ and $u$ bounded to the $[0, 1]$ probability domain. The enforced (equality) constraint is $\int_{l}^{u}f(\mu~|~G_{\mathcal{S}})~d\mu = 1-\alpha$. To accelerate convergence, we use the corresponding $1-\alpha$ \ac{ET} \ac{CrI} as an initial guess for the \ac{SLSQP} optimizer.

\paragraph{\textbf{Limiting Cases.}} The minimality and uniqueness properties of the $1-\alpha$ \ac{HPD} interval can be extended to limiting cases where the beta posterior exhibits exponentially increasing or decreasing behavior. These cases arise when the beta prior is uninformative -- i.e., $a \! = \! b \leq 1$ -- and the annotation process yields either all correct ($\tau_{\mathcal{S}} = n_{\mathcal{S}}$) or all incorrect ($\tau_{\mathcal{S}} = 0$) triples.

When $\tau_{\mathcal{S}} = n_{\mathcal{S}}$, resulting in an exponentially increasing posterior, the $1-\alpha$ \ac{HPD} interval is computed as:
\begin{equation}
    \label{eq:hpdInc}
    l = \operatorname{qBeta}(\alpha; a + \tau_{\mathcal{S}}, b);~u = 1
\end{equation}
Likewise, when $\tau_{\mathcal{S}} = 0$, producing an exponentially decreasing posterior, the $1-\alpha$ \ac{HPD} interval becomes:
\begin{equation}
    \label{eq:hpdDec}
    l = 0;~u = \operatorname{qBeta}(1-\alpha; a, b + n_{\mathcal{S}})
\end{equation}

As a corollary to Theorem~\ref{thm:hpdMin}, we prove that the $1-\alpha$ \ac{HPD} interval is minimal in both limiting cases.

\begin{corollary}
    \label{cor:hpdMin}
    Given an uninformative $\operatorname{Beta}(a, b)$ prior for $\mu$ and a \ac{KG} annotation process resulting in either all correct ($\tau_{\mathcal{S}} = n_{\mathcal{S}}$) or all incorrect ($\tau_{\mathcal{S}} = 0$) triples, the $1-\alpha$ \ac{HPD} interval is the smallest $(l, u)$ interval satisfying $F(u)-F(l) = 1-\alpha$. 
\end{corollary}

\begin{proof}
    We prove $\tau_{\mathcal{S}} = 0$, the proof for $\tau_{\mathcal{S}} = n_{\mathcal{S}}$ is symmetrical. 
    With an uninformative prior (i.e., $a = b \leq 1$) and an annotation outcome $\tau_{\mathcal{S}} = 0$, the resulting beta posterior has parameters $a \leq 1$ and $b + n_{\mathcal{S}} > 1$, exhibiting an exponentially decreasing behavior with the highest density in $x=0$. Due to the monotonicity of the beta posterior, for any $x_{1} < x_{2}$, we have $f(x_{1}~|~G_{\mathcal{S}}) > f(x_{2}~|~G_{\mathcal{S}})$. Thus, the \ac{HPD} interval, with a lower bound at $0$ and an upper bound at the $1-\alpha$ quantile, is by construction the shortest interval containing $\mu$ with probability $1-\alpha$.
\end{proof}

Similarly, we prove its uniqueness as a corollary to Theorem~\ref{thm:hpdUnique}.

\begin{corollary}
    \label{cor:hpdUnique}
    Under the assumptions of Corollary~\ref{cor:hpdMin}, the $1-\alpha$ \ac{HPD} interval is unique.
\end{corollary}

\begin{proof}[Sketch Proof]
    The proof follows the same rationale as the proof of Theorem~\ref{thm:hpdUnique}. We prove for both limiting cases -- $\tau_{\mathcal{S}} = 0$ and $\tau_{\mathcal{S}} = n_{\mathcal{S}}$ -- by contradiction. Suppose another interval $(l', u')$ exists with the same properties as the $1-\alpha$ \ac{HPD} interval. This interval could either overlap with the \ac{HPD} or lie entirely outside its range. In the overlapping case, the proof mirrors that of Theorem~\ref{thm:hpdUnique}. For the external case, due to the monotonicity of the posterior distribution, it is trivial to show that the $(l', u')$ interval cannot have the same width as the \ac{HPD} while covering $\mu$ with probability $1-\alpha$, thus proving the uniqueness of the \ac{HPD} interval.
\end{proof}

\paragraph{\textbf{Posterior Skewness.}} The properties of $1-\alpha$ \ac{HPD} \acp{CrI} make them suited for representing skewed posterior distributions. To verify this, we revisit Figure~\ref{fig:etVShpd} with a focus on \ac{HPD} intervals. When the distribution is skewed, as shown in Figures~\ref{fig:etVShpd}(b,c), the $1-\alpha$ \ac{HPD} interval exhibits two expected properties. First, it is smaller than the corresponding $1-\alpha$ \ac{ET} interval. Secondly, it only covers the most probable parameter values, resulting in a shift compared to the \ac{ET} interval. Thus, for skewed posteriors, \ac{HPD} \acp{CrI} provide the most precise and reliable summaries for post-data inference.  
 
On the other hand, when the posterior is unimodal and symmetric, as in Figure~\ref{fig:etVShpd}(a), the \ac{HPD} interval coincides with the \ac{ET} interval. This relationship is formalized in the following theorem.

\begin{theorem}
    \label{thm:hpd=et}
     When the beta posterior for $\mu$ is unimodal and symmetric, the $1-\alpha$ \ac{HPD} interval is equivalent to the $1-\alpha$ \ac{ET} interval.
\end{theorem}

\begin{proof}[\rtwo{Sketch Proof}]
    \rtwo{Let $\omega$ represent the mode of a unimodal and symmetric posterior. By the symmetry property, for any $\delta > 0$, we have $f(\omega - \delta~|~G_{\mathcal{S}}) = f(\omega + \delta~|~G_{\mathcal{S}})$. Therefore, the $1-\alpha$ \ac{HPD} interval is symmetric around $\omega$ -- i.e., the lower and upper bounds, $l$ and $u$, are equidistant from $\omega$. Consequently, it follows that $\int_{0}^{l}f(\mu~|~G_{\mathcal{S}})~d\mu = \int_{u}^{1}f(\mu~|~G_{\mathcal{S}})~d\mu = \frac{\alpha}{2}$, which correspond to the \acp{CDF} of the $\alpha/2$ and $1-\alpha/2$ quantiles, respectively. That is, the $1-\alpha$ \ac{HPD} and \ac{ET} intervals coincide.}
\end{proof}

This scenario occurs when, given a $\operatorname{Beta}(a, b)$ prior, the annotation process yields a $\operatorname{Beta}(a + \tau_{\mathcal{S}}, b + n_{\mathcal{S}} - \tau_{\mathcal{S}})$ posterior where $a + \tau_{\mathcal{S}} = b + n_{\mathcal{S}} - \tau_{\mathcal{S}}$, which implies a post-data balance between correct and incorrect triples and hence a \ac{KG} accuracy of $0.5$.

\paragraph{\textbf{\rometa{Implications}}}

\rometa{These theoretical results encompass all the scenarios that can be encountered in the annotation process for evaluating \ac{KG} accuracy. These results stem from modeling the task, for the first time, through a beta-binomial process and account for configurations involving informative and uninformative priors. This led to new proofs tailored to the specific requirements of the considered task.}

\rometa{These findings have notable implications for data quality management because they show that \ac{HPD} \acp{CrI} consistently optimize the convergence rate of the \ac{KG} accuracy minimization problem across all possible scenarios. This translates into minimizing the number of annotations required to audit \ac{KG} accuracy, thereby reducing temporal and monetary costs that are critical factors for data curation~\cite{paulheim-2018}.}

\rometa{Furthermore, these theoretical insights extend beyond \ac{KG} accuracy evaluation to any task involving a similar annotation process. As such, they hold broader applicability and significance, offering valuable contributions also to related domains.}

\subsection{The Challenge of Choosing the Right Prior}
\label{subsec:priorSelection}

The Bayesian framework we designed for \ac{KG} accuracy evaluation offers a seamless approach to build an infinite number of $1-\alpha$ \acp{CrI} by varying the $a, b$ parameters of the beta prior. This contrasts with frequentist methods, which involve different and complex procedures to generate $1-\alpha$ \acp{CI} that are challenging to interpret~\cite{makowski_etal-2019,morey_etal-2016}. However, adopting the unified and flexible Bayesian approach requires choosing appropriate priors that optimize efficiency while ensuring reliability.

We focus on the most general and challenging scenario, where no prior knowledge or belief about the \ac{KG} accuracy is available. In this case, using informative (or subjective) priors can be detrimental, leading to deceptive interpretations or inefficient solutions~\cite{lesaffre_lawson-2012}. To mitigate the bias introduced by these priors, we can resort to uninformative (or objective) priors~\cite{press-2002}, which reflect a lack of available information about the underlying process. In other words, uninformative priors do not encode the analyst's particular belief (or bias), making them suited to provide an objective analytical foundation. This allows the framework to be used by anyone -- layman or expert -- regardless of their specific expectations about the outcomes of the \ac{KG} evaluation process. 

For beta distributions, uninformative priors are characterized by parameters $a=b\leq 1$. In the literature, three proper (i.e., $a,b > 0$) uninformative beta priors are widely used: Kerman $\operatorname{Beta}(\frac{1}{3}, \frac{1}{3})$~\cite{kerman-2011}, Jeffreys $\operatorname{Beta}(\frac{1}{2}, \frac{1}{2})$~\cite{jeffreys-1946}, and Uniform $\operatorname{Beta}(1, 1)$~\cite{bayes-1763}. The Jeffreys prior is commonly adopted as the default choice for binomial proportion problems~\cite{brown_etal-2001}. However, it represents a trade-off between Kerman and Uniform priors~\cite{jin_etal-2017}, and as a result, it is never the most efficient option. 

\begin{figure}[t!]
\centering
\includegraphics[width=0.9\linewidth]{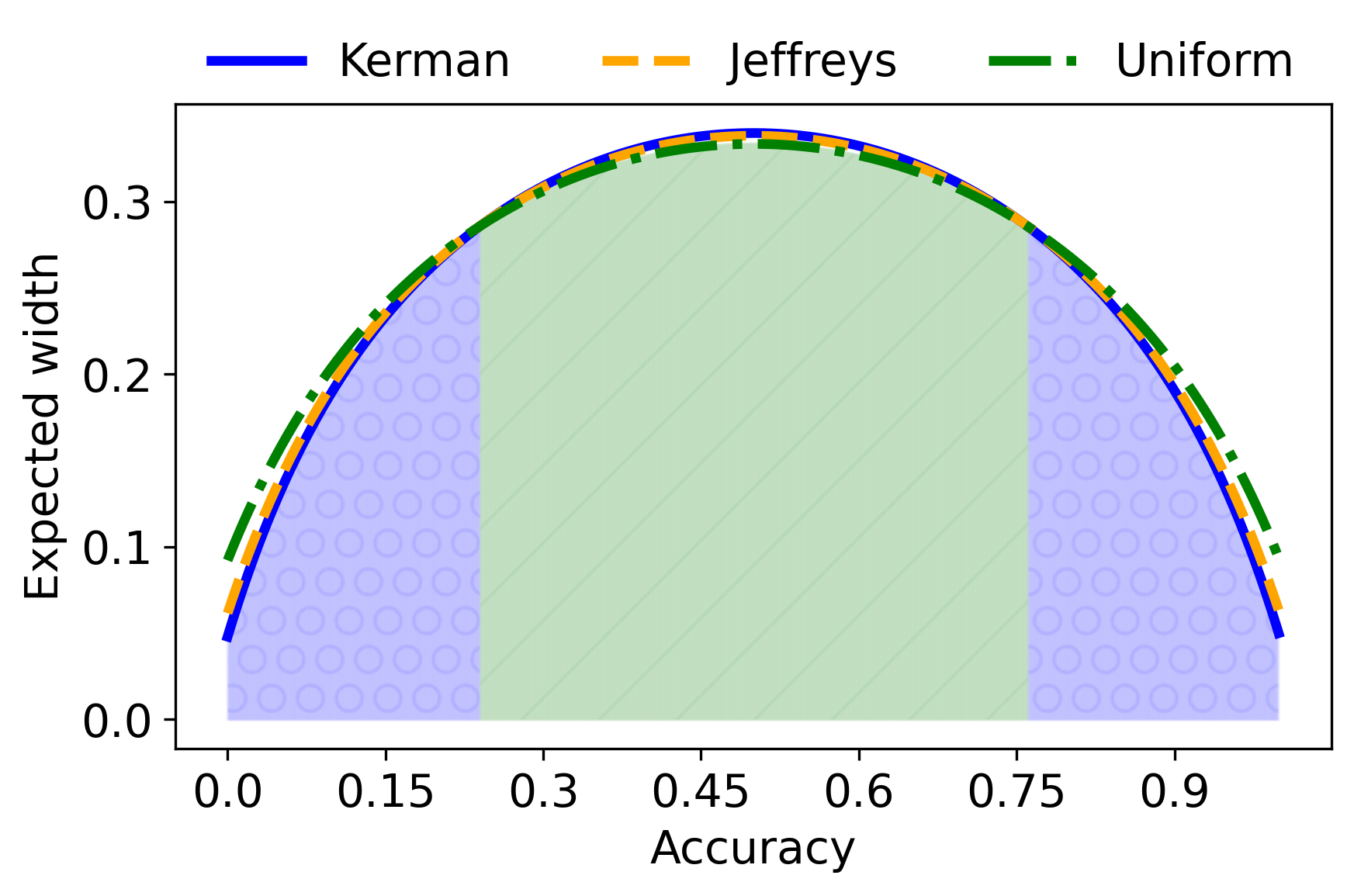}
  \caption{Expected width of HPD credible intervals under Kerman, Jeffreys, and Uniform priors for $n_{\mathcal{S}} = 30$ and $\alpha = 0.05$. The circle pattern ($\circ$) under the curves represents the set of accuracy values where Kerman prior provides the shortest expected width, whereas the line pattern ($//$) the set of accuracy values where Uniform performs best.}
  \label{fig:priorWidth}
\end{figure}

As an example, Figure~\ref{fig:priorWidth} presents the expected width of $1-\alpha$ \ac{HPD} intervals under Kerman, Jeffreys, and Uniform priors for $n_{\mathcal{S}}=30$ and $\alpha=0.05$. We can see that the \ac{HPD} interval based on Jeffreys prior is never the shortest of the three. In contrast, Kerman- and Uniform-based \ac{HPD} intervals achieve minimal widths depending on the specific accuracy space regions. Kerman-based intervals are optimal in the extreme regions, while Uniform-based intervals are optimal in the central region. 

\rometa{Despite being minor, these differences in interval widths lead to} \rometa{appreciable variations in annotation costs for \ac{KG} accuracy evaluation. As we will see experimentally in Section~\ref{subsec:priorExps}, the annotation costs obtained under different priors are statistically different in most cases. This further underscores the importance of addressing the prior selection problem.}

Therefore, choosing the right uninformative prior is a nontrivial decision. Since the outcome of the annotation process -- and thus the region of the accuracy space where the estimate will fall -- cannot be known in advance, it is impossible to choose, \emph{a priori}, one prior that consistently optimizes efficiency. 

\subsection{The \textit{adaptive} HPD (\emph{a}HPD) algorithm}
\label{subsec:aHPD}

To address the challenge of prior selection, we propose the \acf{aHPD} algorithm (see Algorithm~\ref{alg:aHPD}), where multiple uninformative priors are concurrently employed to generate distinct $1-\alpha$ \ac{HPD} intervals. 

\begin{algorithm}[t!]
\centering
\caption{\textit{adaptive} HPD}\label{alg:aHPD}
\begin{algorithmic}[1]
\Require 
\Statex The KG $G$;
\Statex The significance level $\alpha$;
\Statex The sampling strategy $\mathcal{S}$;
\Statex The upper bound $\varepsilon$ for the MoE;
\Statex The set of uninformative beta prior parameters $\{(a_{1}, b_{1}), \ldots, (a_{k}, b_{k})\}$;
\Ensure The estimated accuracy and the smallest $1-\alpha$ HPD interval.

\State Define objective function (\texttt{obj}) to minimize interval width $(u - l)$
\State Define constraint function (\texttt{constr}) to satisfy $F(u) - F(l) = 1 - \alpha$
\State $\texttt{bounds} \leftarrow [[0, 1], [0, 1]]$ \Comment{Set $l$ and $u$ bounds}
\State $\texttt{mu} \leftarrow 0$, $\texttt{moe} \leftarrow 1$, $\texttt{sample} \leftarrow \emptyset$ \Comment{Initialization}
\While{$\texttt{moe} > \varepsilon$}
\State $\texttt{batch} \leftarrow \mathcal{S}(G)$ \Comment{Sample batch of facts from $G$ via $\mathcal{S}$}
\State $\texttt{annot} \leftarrow \mathds{1}(\texttt{batch})$ \Comment{Annotate batch and store annotations}
\State $\texttt{sample} \leftarrow \texttt{sample} \cup \texttt{annot}$ \Comment{Update sample with new annotations} 
\State $\texttt{mu} \leftarrow \hat{\mu}_{\mathcal{S}}(\texttt{sample})$ \Comment{Estimate $G$ accuracy from sample}
\State $\texttt{tau} \leftarrow \operatorname{sum}(\texttt{annot})$, $\texttt{n} \leftarrow |\texttt{sample}|$ \Comment{Correct triples and sample size}
\If{$\mathcal{S} \; ! \! = \, \text{SRS}$} \Comment{Complex sampling strategy}
\State Adjust for complex sampling effects using design effect corrections~\cite{kish-1995,marchesin_silvello-2024}
\EndIf
\State Compute posterior parameters from each prior and given annotation outcome: 
\Statex $\; \; \; \; \; (\texttt{a\_post}_{i}, \texttt{b\_post}_{i}) \leftarrow (a_{i} + \texttt{tau}, b_{i} + \texttt{n}-\texttt{tau})$ 
\If{$\texttt{tau} == \texttt{n}$} \Comment{Limiting case (1): all facts correct} 
\State $\texttt{l}_{i} \leftarrow \operatorname{qBeta}(\alpha; \texttt{a\_post}_{i}, \texttt{b\_post}_{i})$, $\texttt{u}_{i} \leftarrow 1 \, \forall i \in \{1, \ldots, k\}$ 
\ElsIf{$\texttt{tau} == 0$} \Comment{Limiting case (2): no correct facts}
\State $\texttt{l}_{i} \leftarrow 0$, $\texttt{u}_{i} \leftarrow \operatorname{qBeta}(1-\alpha; \texttt{a\_post}_{i}, \texttt{b\_post}_{i}) \, \forall i \in \{1, \ldots, k\}$ 
\Else \Comment{Standard case: unimodal posterior}
\State Compute, for each posterior, the $1-\alpha$ ET CrI as initial guess:
\Statex $\; \; \; \; \; \; \; \; \; \, \, \texttt{guess}_{i} \leftarrow \operatorname{ET}(1-\alpha; \texttt{a\_post}_{i}, \texttt{b\_post}_{i})$ 
\State Compute, for each posterior, the $1-\alpha$ HPD CrI using SLSQP optimizer:
\Statex $\; \; \; \; \; \; \; \; \; \, \, (\texttt{l}_{i}, \texttt{u}_{i}) \leftarrow \text{SLSQP}(\texttt{obj}, \texttt{guess}_{i}, \texttt{constr}, \texttt{bounds})$ 
\EndIf
\State $(\texttt{l}, \texttt{u}) \leftarrow \operatorname{arg\,min}_{i}(\texttt{u}_{i}-\texttt{l}_{i})$ \Comment{Select smallest $1-\alpha$ HPD CrI}
\State $\texttt{moe} \leftarrow (\texttt{u}-\texttt{l})/2$ \Comment{Update MoE}
\EndWhile
\State \Return \texttt{mu}, $(\texttt{l}, \texttt{u})$
\end{algorithmic}
\end{algorithm}

At each stage of the \ac{KG} evaluation process, the algorithm samples and annotates a batch of facts (lines 6-7), which combines with previous annotations to update the accuracy estimate (lines 8-9). Based on the number of correct facts and the sample size, \ac{aHPD} first applies design effect adjustments in case of complex sampling strategies~\cite{kish-1995,marchesin_silvello-2024} and then calculates the posterior parameters for each uninformative prior provided as input (lines 10-14); there is no limit to the number of priors that can be fed to the algorithm. It then builds the corresponding $1-\alpha$ \ac{HPD} intervals. When the number of correct facts equals the sample size (lines 15-16), in the first limiting case, the posterior distribution takes an exponentially increasing shape, and the interval bounds are computed using Equation~(\ref{eq:hpdInc}). When the number of correct facts is zero (lines 17-18), in the second limiting case, the posterior distribution follows an exponentially decreasing pattern, and the intervals are derived via Equation~(\ref{eq:hpdDec}). For the standard case, having defined the objective function, constraints, and bounds (lines 1-3), the algorithm adopts the SLSQP optimizer to generate the $1-\alpha$ \ac{HPD} intervals, using the corresponding \ac{ET} intervals as initial guesses (lines 20-21). It selects the smallest one among the generated intervals and computes its \ac{MoE} (lines 23-24). The process iterates until $\text{MoE} \leq \varepsilon$ (line 5), ensuring the selected $1-\alpha$ \ac{HPD} interval meets the desired precision. Once the criterion is satisfied, the algorithm returns the estimated accuracy and the chosen interval. Note that the computation of posterior parameters (line 14) and the construction of $1-\alpha$ intervals (lines 16, 18, 20-21) can be parallelized for each input prior, ensuring that the algorithm remains efficient regardless of the number of considered priors. 

The \ac{aHPD} algorithm dynamically adjusts to the region of the accuracy space where the estimate lies after the annotation process. This adaptive approach ensures that the first interval achieving the desired precision halts the evaluation process, guaranteeing the most efficient outcome from the set of competing solutions.

\paragraph{\textbf{\rometa{aHPD with Informative Priors}}}

\rometa{Although in this work we focus on the most general and challenging scenario, where no prior knowledge or belief about the KG accuracy is available, if we have (reliable) prior knowledge, then \ac{aHPD} can be seamlessly used with such informative priors to optimize performance further.}
\begin{example}
\label{ex:infPrior}
    \rometa{Let us assume that an analyst who wants to estimate the accuracy of DBpedia ($\mu=0.85$, see Table~\ref{tab:kgStats}) already knows the accuracy of two similar \acp{KG} -- which have accuracy values $\mu=0.80$ and $\mu=0.90$, respectively. Based on this knowledge, the analyst can set up two informative priors, e.g., ($a_1=80$, $b_1=20$) and ($a_2=90$, $b_2=10$), and plug them into \ac{aHPD}. The performance obtained with these informative priors under \ac{TWCS}, averaged over $1,000$ repetitions, are $63\pm36$ triples and a cost of $0.72\pm0.41$, instead of $222\pm83$ triples and a cost of $2.55\pm0.95$ when using \ac{aHPD} equipped with Kerman, Jeffreys, and Uniform uninformative priors.}
\end{example}
\rometa{Hence, \ac{aHPD} can lead to even more efficient solutions if prior knowledge is available. Meanwhile, poor informative priors can be detrimental, leading to deceptive interpretations or inefficient solutions. In these cases, relying on uninformative priors is better.}

\section{Experimental Setup}
\label{sec:setup}

In the following, we report the considered datasets, the adopted cost function, the experimental configuration, and the evaluation procedure. We release data and code used in this work.\footnote{\url{https://github.com/KGAccuracyEval/credible-intervals4kg-estimation}} Moreover, we provide more experimental results presenting additional sampling strategies in the online repository. While these results are consistent with those given in the main text, they are less critical and thus are made available online due to space limits.

\paragraph{\textbf{Datasets}}
\label{subsec:datasets}

We consider data from various real-life \acp{KG}, whose statistics are reported in Table~\ref{tab:kgStats}. \rtwo{Note that dataset sizes are rather small due to the manual labeling costs.}

\begin{table}[t!]
\caption{Statistics for YAGO, NELL, DBPEDIA, FACTBENCH, and SYN 100M datasets.}
\centering
\begin{tabular}{lcccccc}
\toprule
                      & \multicolumn{1}{c}{YAGO} & \multicolumn{1}{c}{NELL} & \multicolumn{1}{c}{DBPEDIA} & \multicolumn{1}{c}{FACTBENCH} & \multicolumn{1}{c}{SYN $100$M} \\ \midrule
Num. of facts       & 1,386                    & 1,860                    & 9,344 & 2,800 & 101,415,011            \\
Num. of clusters    & 822                      & 817                      & 2,936 & 1,157 &  5,000,000             \\
Avg. cluster size          & 1.69            & 2.28            & 3.18 & 2.42 & 20.28    \\
Accuracy ($\mu$) & 0.99                     & 0.91                     & 0.85 & 0.54 & \{0.9,0.5,0.1\}         \\ \bottomrule
\end{tabular}
\label{tab:kgStats}
\end{table}

\textbf{YAGO}: a dataset sampled by \citet{ojha_talukdar-2017} from the YAGO2 KG~\cite{hoffart_etal-2013}, with facts referring to people, organizations, countries, and movies. The dataset was manually annotated \rtwo{by crowdsourcing workers}, resulting in a ground-truth accuracy of $\mu = 0.99$.

\textbf{NELL}: drawn by \citet{ojha_talukdar-2017} from the NELL \ac{KG}~\cite{mitchell_etal-2018}, this dataset focuses on sports-related facts. It includes manual annotations \rtwo{by crowdsourcing workers} for each fact, yielding a ground-truth accuracy of $\mu = 0.91$.

\textbf{DBPEDIA}: a dataset sampled by \citet{marchesin_etal-2024b} from DBpedia~\cite{auer_etal-2007},\footnote{\url{https://downloads.dbpedia.org/wiki-archive/dbpedia-dataset-version-2015-10.html}} covering a broad range of topics such as entertainment, news, history, sports, business, and science. \rtwo{Each fact in the dataset was manually annotated by at least three layman workers. The correctness label for every fact was then derived via quality-weighted majority voting, where (layman) worker quality was evaluated on an extra pool of facts supervised by expert workers. This annotation process resulted in a ground-truth accuracy of $\mu = 0.85$.}

\textbf{FACTBENCH}: a benchmark created by \citet{gerber_etal-2015} to assess fact validation algorithms. It contains correct facts from DBpedia~\cite{auer_etal-2007} and Freebase~\cite{bollacker_etal-2007,bollacker_etal-2008}, and generates incorrect facts by modifying correct ones while adhering to domain and range constraints. We consider a configuration where incorrect facts are a mix produced using different negative example generation strategies, resulting in a ground-truth accuracy of $\mu = 0.54$. 

YAGO and NELL are the reference datasets used in the literature to evaluate \ac{KG} accuracy estimation approaches~\cite{ojha_talukdar-2017,gao_etal-2019,qi_etal-2022,marchesin_silvello-2024}. The DBPEDIA dataset is a proxy for one of the largest and widely used real-life \acp{KG}. Conversely, FACTBENCH represents a controlled scenario designed to assess the ability of automatic fact validation methods to distinguish between real (correct) and synthetic (incorrect) facts. Although unrealistic, FACTBENCH can be used to evaluate the behavior of \ac{KG} accuracy estimation solutions in the quasi-symmetric case. Note that other datasets used by prior work exist, such as MOVIE~\cite{gao_etal-2019} and OPIEC~\cite{qi_etal-2022}, but they are either not publicly (MOVIE) or only partially (OPIEC) available.

We assess the scalability of the methods by using \textbf{SYN 100M}~\cite{marchesin_silvello-2024}, a synthetic dataset with over $100$ million triples, where correctness labels are generated by setting the probability of a triple being true to a fixed rate. We consider ground-truth accuracy values of $\mu \in \{0.9, 0.5, 0.1\}$.

\paragraph{\textbf{Annotation Cost}}
\label{subsec:annotationCost}

We use the cost function proposed by~\cite{gao_etal-2019} and later adopted in~\cite{qi_etal-2022,marchesin_silvello-2024} to measure the manual effort required to assess the correctness of facts in a given sample. \rthree{The function is based on the assumption that annotating a fact for an entity already identified incurs a lower cost than annotating a fact from an unseen entity. Specifically, it divides the annotation process into entity identification, which involves linking the entity to the corresponding real-world concept, and fact verification, which entails collecting evidence and auditing the fact stated by the triple.} 

\rthree{The cost function is expressed as:}
\begin{equation}
    \label{eq:costFunction}
    \operatorname{cost}(G_{\mathcal{S}}) = |E_{\mathcal{S}}|\cdot c_{1} + |T_{\mathcal{S}}| \cdot c_{2}
\end{equation}
Here, $c_{1}$ and $c_{2}$ represent the average time cost (in seconds) for entity identification and fact verification, respectively. In line with prior research~\cite{gao_etal-2019,qi_etal-2022,marchesin_silvello-2024}, we set $c_{1} = 45$ and $c_{2} = 25$ (seconds).

\paragraph{\textbf{Sampling Strategies}} 
We adopt \ac{SRS} and \ac{TWCS} as sampling strategies (Section~\ref{subsec:sampling}). Based on the recommendation by \citet{gao_etal-2019} to choose the second-stage size $m$ of \ac{TWCS} in the range $\{3, 5\}$, we set $m=3$ for YAGO, NELL, DBPEDIA, and FACTBENCH, which have small clusters, and $m=5$ for SYN 100M, which presents larger clusters (see Table~\ref{tab:kgStats}).

\paragraph{\textbf{Interval Estimation}}
We consider Wald (Section~\ref{subsec:wald}) and Wilson (Section~\ref{subsec:wilson}) \acp{CI} as baselines and compare them with the \ac{aHPD} (Section~\ref{subsec:aHPD}). Wald represents an efficient yet unreliable baseline, which we consider only to emphasize the efficiency of our approach further. Conversely, Wilson represents the current state-of-the-art, providing a trade-off between efficiency and (frequentist) reliability. For \ac{aHPD}, we employ three uninformative priors: Kerman $\operatorname{Beta}(\frac{1}{3}, \frac{1}{3})$, Jeffreys $\operatorname{Beta}(\frac{1}{2}, \frac{1}{2})$, and Uniform $\operatorname{Beta}(1, 1)$. We also evaluate the performance of \ac{ET} (Section~\ref{subsec:etInt}) and \ac{HPD} (Section~\ref{subsec:hpdInt}) \acp{CrI} under these priors. We apply the design effect adjustments from~\cite{marchesin_silvello-2024} when the considered intervals are used with \ac{TWCS}.

\paragraph{\textbf{Evaluation Procedure}}
\label{subsec:evalProc}

We set the significance level at $\alpha \in \{0.10, 0.05, 0.01\}$ and define the upper bound for the \ac{MoE} as $\varepsilon = 0.05$. When not specified, the significance level defaults to $\alpha = 0.05$. We use two metrics to compare methods: the number of annotated triples and the annotation cost (in hours). \rthree{The number of triples represents a simple and plain measure, while the annotation cost reflects a more realistic measure of temporal (and monetary) costs.} 

Results are reported when $\text{MoE} \leq 0.05$, ensuring the methods are compared once they satisfy the optimization constraint. For this reason, we do not report the width of \acp{CI}, as all solutions have $\text{MoE} \leq 0.05$. Similarly, accuracy estimates are omitted since all methods provide unbiased estimates with minimal deviation from ground-truth accuracy (less than $0.02$). 

To account for the inherent variability of sampling strategies, we repeat the evaluation procedure $1,000$ times per method, reporting the mean and standard deviation for both the number of triples and the annotation cost.

\section{Experimental Results}
\label{sec:results}

The objectives of the experiments are to (1) assess the impact of prior selection on \ac{KG} accuracy estimation methods; (2) evaluate the efficiency of the \ac{aHPD} compared to the state-of-the-art; (3) test whether \ac{aHPD} is still efficient with large-scale datasets; and, (4) verify the robustness of \ac{aHPD} across different levels of precision. 

\subsection{Summary of the Experimental Findings}
\label{subsec:findings}

\begin{itemize}
    \item[(F1)] \textit{There is no one-prior-fits-all solution}. Section~\ref{subsec:priorExps} shows that there is no single prior that we can apply to obtain top performance consistently, thus motivating the need for the \ac{aHPD}.
    \item[(F2)] \textit{aHPD outperforms state-of-the-art solutions in all real-life cases}. Section~\ref{subsec:efficiency} shows that \ac{aHPD} reduces convergence costs compared to Wald and Wilson in common cases with a skewed accuracy distribution.
    \item[(F3)] \textit{aHPD scales efficiently to large datasets.} Section~\ref{subsec:scalability} shows that the dataset size does not affect the performance of \ac{aHPD}, which remains the most efficient solution. 
    \item[(F4)] \textit{aHPD is robust across varying levels of precision}. Section~\ref{subsec:robustness} shows that \ac{aHPD} preserves its optimal performance regardless of the precision level required by the evaluation process.
\end{itemize}

\subsection{Prior Selection}
\label{subsec:priorExps}

\rone{The performance of \ac{ET} and \ac{HPD} \acp{CrI} under Kerman, Jeffreys, and Uniform priors are reported in Table~\ref{tab:priorComp}, together with those of \ac{aHPD} equipped with the three considered priors}. For sampling, we resorted to \ac{SRS}, which allows for a clean comparison between \acp{CrI} built from different priors without introducing clustering. Such effects can hamper the interpretation of the results, making it challenging to determine whether the observed outcomes are due to the specific prior under consideration or to clustering effects.

\begin{table}[t!]

\caption{Performance of \ac{ET} and \ac{HPD} \acp{CrI} on YAGO, NELL, DBPEDIA, and FACTBENCH under different priors using SRS. For each interval, the best performances are in bold. \rone{We also report the performance obtained by \ac{aHPD} equipped with the considered priors.}}
\begin{tabular}{llccccccc}
\toprule
\multicolumn{2}{c}{} & \multicolumn{1}{c|}{YAGO}                                                 & \multicolumn{1}{c|}{NELL}                                                  & \multicolumn{1}{c|}{DBPEDIA} & \multicolumn{1}{c}{FACTBENCH} \\ \midrule
   \multicolumn{2}{c}{}                  & \multicolumn{1}{c|}{$\mu = 0.99$}                                           & \multicolumn{1}{c|}{$\mu = 0.91$}                                            & \multicolumn{1}{c|}{$\mu = 0.85$} & \multicolumn{1}{c}{$\mu = 0.54$}                           \\ \midrule
Interval & Prior               & \multicolumn{1}{c|}{Triples}              & \multicolumn{1}{c|}{Triples}    &  \multicolumn{1}{c|}{Triples}     & \multicolumn{1}{c}{Triples} \\ \midrule
\multirow{3}{*}{ET} & Kerman & $\mathbf{38\!\pm\!6}$ & $\mathbf{104\!\pm\!40}$ & $\mathbf{187\!\pm\!36}$ & $380\!\pm\!3$	  \\ 
& Jeffreys	& $39\!\pm\!7$  & $107\!\pm\!39$ & $188\!\pm\!34$ & $379\!\pm\!3$ \\ 
& Uniform	& $41\!\pm\!9$ & $113\!\pm\!36$ & $190\!\pm\!32$ & $\mathbf{378\!\pm\!3}$ \\ 
\midrule
\multirow{3}{*}{HPD} & Kerman & $\mathbf{32\!\pm\!5}$ & $\mathbf{96\!\pm\!44}$ & $\mathbf{182\!\pm\!42}$ & $380\!\pm\!3$	  \\ 
& Jeffreys & $33\!\pm\!6$ & $99\!\pm\!42$ & $184\!\pm\!39$ & $379\!\pm\!3$	  \\ 
& Uniform & $34\!\pm\!9$ & $106\!\pm\!40$ & $187\!\pm\!36$ & $\mathbf{378\!\pm\!3}$	  \\  \midrule
\rone{\ac{aHPD}} & \rone{$\{$K, J, U$\}$} & \rone{$\mathbf{32\!\pm\!5}$}	& \rone{$\mathbf{96\!\pm\!44}$} & \rone{$\mathbf{182\!\pm\!42}$} & \rone{$\mathbf{378\!\pm\!3}$} \\ 
\bottomrule
\end{tabular}
\label{tab:priorComp}
\end{table}

The results in Table~\ref{tab:priorComp} confirm that Kerman and Uniform priors lead to the most efficient solutions in the extreme and central regions of the accuracy space, respectively (cf. Section~\ref{subsec:priorSelection}). Conversely, the Jeffreys prior never achieves top performance. This pattern holds for both \ac{ET} and \ac{HPD} \acp{CrI}, highlighting the generality of the problem of choosing an appropriate prior.

\rone{Focusing on \ac{HPD} intervals, the annotation costs under different priors are statistically different in 8 out of 12 cases, as determined by standard independent t-tests with $p < 0.01$. This emphasizes that the choice of the prior significantly affects the outcomes. In this regard, \ac{aHPD}, which automatically selects the optimal prior (cf. Table~\ref{tab:priorComp}), proves to be statistically better than \ac{HPD} solutions relying on suboptimal priors in $75\%$ of cases. Therefore, \ac{aHPD} emerges as a robust approach for consistently selecting the optimal prior for a given annotation process, removing the need to choose \emph{a priori}.}

It is worth noting that other priors could be explored, which may perform better than the considered three in specific regions of the accuracy space. This further stresses the problem of prior selection when optimizing efficiency. For this work, we focused on three of the most popular priors, showing that even with a relatively small set, it is impossible to guarantee optimal performance across the entire accuracy space without knowing where the accuracy estimate will fall after the annotation process. This strengthens the need for an adaptive solution as the \ac{aHPD}.

As a side note, Table~\ref{tab:priorComp} also shows that \ac{HPD} intervals are better than \ac{ET} ones when the accuracy is skewed (YAGO, NELL, DBPEDIA) and equal when the accuracy is quasi-symmetric (FACTBENCH), as proved by Theorems~\ref{thm:hpdMin}-\ref{thm:hpd=et}.

\subsection{Efficiency}
\label{subsec:efficiency}

The performance of \ac{aHPD} against Wald and Wilson baselines is reported in Table~\ref{tab:efficiency}. We compared methods considering both \ac{SRS} and \ac{TWCS} as sampling strategies. \rttmeta{We also performed statistical analyses to assess the significance of the performance differences between \ac{aHPD} and baselines. Specifically, we conducted standard independent t-tests between \ac{aHPD} and Wald, and between \ac{aHPD} and Wilson, reporting statistical results when $p < 0.01$.}

\begin{table}[t!]

\caption{Performance on YAGO, NELL, DBPEDIA, and FACTBENCH. For TWCS, we set $m=3$. For each sampling strategy, best performance are in bold. \rttmeta{$^{\dagger}$ denotes a statistically significant difference between \ac{aHPD} and Wald, while $^{\ddagger}$ indicates a significant difference between \ac{aHPD} and Wilson, as determined by standard independent t-tests with $p < 0.01$.}}
\scriptsize
\begin{tabular}{llcccccccc}
\toprule
\multicolumn{2}{c}{} & \multicolumn{2}{c|}{YAGO}                                                 & \multicolumn{2}{c|}{NELL}                                                  & \multicolumn{2}{c|}{DBPEDIA} & \multicolumn{2}{c}{FACTBENCH} \\ \midrule
   \multicolumn{2}{c}{}                  & \multicolumn{2}{c|}{$\mu = 0.99$}                                           & \multicolumn{2}{c|}{$\mu = 0.91$}                                            & \multicolumn{2}{c|}{$\mu = 0.85$} & \multicolumn{2}{c}{$\mu = 0.54$}                           \\ \midrule
Sampling & Interval               & Triples   & \multicolumn{1}{c|}{Cost}           & Triples    & \multicolumn{1}{c|}{Cost}          & Triples     & \multicolumn{1}{c|}{Cost} & Triples     & \multicolumn{1}{c}{Cost} \\ \midrule
\multirow{3}{*}{SRS} & Wald	 & $33\!\pm\!7$	 & $0.62\!\pm\!0.12$	 & $103\!\pm\!43$	 & $1.89\!\pm\!0.76$	 & $188\!\pm\!38$	 & $3.55\!\pm\!0.71$ & $382\!\pm\!3$ & $6.37\!\pm\!0.10$	 \\ 
& Wilson	 & $41\!\pm\!10$	 & $0.76\!\pm\!0.18$	 & $114\!\pm\!36$	 & $2.08\!\pm\!0.63$	 &  $190\!\pm\!32$	 & $3.60\!\pm\!0.60$ & $\mathbf{378\!\pm\!3}$ & $\mathbf{6.32\!\pm\!0.10}$	 \\ 
\rowcolor{purple} & \emph{a}HPD	 & $\mathbf{32\!\pm\!5}$	 & $\mathbf{0.60\!\pm\!0.09}^{\dagger,\ddagger}$ & $\mathbf{96\!\pm\!44}$	 & $\mathbf{1.76\!\pm\!0.79}^{\dagger,\ddagger}$ & $\mathbf{182\!\pm\!42}$	 & $\mathbf{3.45\!\pm\!0.78}^{\dagger,\ddagger}$  & $\mathbf{378\!\pm\!3}$ & $\mathbf{6.32\!\pm\!0.10}^{\dagger}$ \\
\midrule
\multirow{3}{*}{TWCS} & Wald	 & $32\!\pm\!4$	 & $0.42\!\pm\!0.05$ & $126\!\pm\!70$	 & $1.58\!\pm\!0.87$ & $243\!\pm\!75$	 & $2.80\!\pm\!0.87$ & $\mathbf{254\!\pm\!37}$ & $\mathbf{3.08\!\pm\!0.45}$ \\ 
& Wilson	 & $35\!\pm\!4$	 & $0.47\!\pm\!0.06$ & $129\!\pm\!67$	 & $1.62\!\pm\!0.84$ & $234\!\pm\!73$	 & $2.69\!\pm\!0.84$	& $257\!\pm\!39$ & $3.11\!\pm\!0.48$ \\ 
\rowcolor{purple} & \emph{a}HPD	 & $\mathbf{31\!\pm\!2}$	 & $\mathbf{0.41\!\pm\!0.03}^{\dagger,\ddagger}$ & $\mathbf{112\!\pm\!68}$	 & $\mathbf{1.40\!\pm\!0.85}^{\dagger,\ddagger}$ & $\mathbf{222\!\pm\!83}$	 & $\mathbf{2.55\!\pm\!0.95}^{\dagger,\ddagger}$ & $257\!\pm\!39$ & $3.11\!\pm\!0.48$ \\  
\bottomrule
\end{tabular}
\label{tab:efficiency}
\end{table}

Table~\ref{tab:efficiency} provides the following insights. \ac{aHPD} \rttmeta{statistically} outperforms both Wald and Wilson methods across all datasets where the \ac{KG} accuracy is skewed -- i.e., YAGO, NELL, and DBPEDIA. This is a significant outcome, as it shows that \ac{aHPD} surpasses not only the Wilson method, which sacrifices efficiency to ensure high (frequentist) reliability~\cite{marchesin_silvello-2024} but also the unreliable, yet very efficient Wald method. Thus, \ac{aHPD}, grounded in Bayesian statistics, provides interpretable and reliable results while improving efficiency.

For datasets with quasi-symmetric accuracy distributions, such as FACTBENCH, the performance of \ac{aHPD} matches those of the Wilson method. This result is expected. Since FACTBENCH has an accuracy of $\mu=0.54$, we know that, among the considered priors, the Uniform one achieves optimal performance (see Table~\ref{tab:priorComp}). We also know that, in this scenario, \ac{HPD} and \ac{ET} intervals are the same (Theorem~\ref{thm:hpd=et}). Then, since the Wilson \ac{CI} can be seen as an approximation of the \ac{ET} \ac{CrI} based on the Uniform prior~\cite{jin_etal-2017}, it follows that \ac{aHPD} and Wilson-based solutions yield equal performance. Hence, these findings indicate that \ac{aHPD} offers an efficient and reliable solution suited to all the considered scenarios. It consistently provides \rttmeta{statistically} superior performance in cases of skewed \ac{KG} accuracy while remaining competitive in (quasi-)symmetric situations.

\subsection{Scalability}
\label{subsec:scalability}

\begin{table}[t!]

\caption{Performance on SYN 100M with accuracy values $\mu \in \{0.9, 0.5, 0.1\}$. For TWCS we set $m=5$. For each sampling strategy, best performance are in bold. \rttmeta{$^{\dagger}$ denotes a statistically significant difference between \ac{aHPD} and Wald, while $^{\ddagger}$ indicates a significant difference between \ac{aHPD} and Wilson, as determined by standard independent t-tests with $p < 0.01$.}}
\begin{tabular}{llcccccc}
\toprule
   \multicolumn{2}{c}{}    & \multicolumn{6}{c}{SYN $100$M} \\ \midrule
   \multicolumn{2}{c}{}    & \multicolumn{2}{c|}{$\mu = 0.9$}                                                    & \multicolumn{2}{c|}{$\mu = 0.5$}                                                   & \multicolumn{2}{c}{$\mu = 0.1$}                                      \\ \midrule
Sampling & Interval & Triples      & \multicolumn{1}{c|}{Cost}             & Triples      & \multicolumn{1}{c|}{Cost}            & Triples      & \multicolumn{1}{c}{Cost}  \\ \midrule
\multirow{3}{*}{SRS} & Wald & $122\!\pm\!43$	 & $2.37\!\pm\!0.83$ & $384\!\pm\!1$	 & $7.46\!\pm\!0.03$ & $124\!\pm\!43$	 & $2.42\!\pm\!0.83$   \\
& Wilson & $131\!\pm\!34$	 & $2.54\!\pm\!0.66$ & $\mathbf{380\!\pm\!1}$	 & $\mathbf{7.39\!\pm\!0.03}$ & $133\!\pm\!35$	 & $2.58\!\pm\!0.68$ \\ 
\rowcolor{purple} & \emph{a}HPD & $\mathbf{114\!\pm\!46}$	 & $\mathbf{2.22\!\pm\!0.89}^{\dagger,\ddagger}$ & $\mathbf{380\!\pm\!1}$	 & $\mathbf{7.39\!\pm\!0.03}^{\dagger}$ & $\mathbf{117\!\pm\!45}$	 & $\mathbf{2.28\!\pm\!0.88}^{\dagger,\ddagger}$ \\ \midrule
\multirow{3}{*}{TWCS} & Wald & $120\!\pm\!50$	 & $1.13\!\pm\!0.48$ & $384\!\pm\!63$	 & $3.64\!\pm\!0.60$  & $121\!\pm\!53$	 & $1.14\!\pm\!0.50$ \\
& Wilson & $121\!\pm\!47$	 & $1.14\!\pm\!0.45$ & $\mathbf{374\!\pm\!65}$	 & $\mathbf{3.54\!\pm\!0.62}$ & $121\!\pm\!49$	 & $1.15\!\pm\!0.47$ \\ 
\rowcolor{purple} & \emph{a}HPD & $\mathbf{106\!\pm\!52}$	 & $\mathbf{1.01\!\pm\!0.49}^{\dagger,\ddagger}$  & $\mathbf{374\!\pm\!65}$	 & $\mathbf{3.54\!\pm\!0.62}^{\dagger}$ & $\mathbf{108\!\pm\!54}$	 & $\mathbf{1.02\!\pm\!0.51}^{\dagger,\ddagger}$ \\ 
\bottomrule
\end{tabular}
\label{tab:scalability}
\end{table}

Table~\ref{tab:scalability} reports the \ac{aHPD} performance on SYN 100M datasets compared to the Wald and Wilson methods using \ac{SRS} and \ac{TWCS} for sampling. \rttmeta{As before, we conducted statistical analyses to assess the significance of \ac{aHPD} performance with respect to baselines.} 

We can see that the performance of all methods remain within the same order of magnitude as those observed in small-scale datasets (cf. Table~\ref{tab:efficiency}). This suggests that \ac{KG} accuracy estimation methods are not affected by the size of the datasets. Such a finding is unsurprising, as both the accuracy estimators and the $1-\alpha$ intervals are independent of the underlying (dataset) population. In the case of \acp{CI}, they only rely on the sample and its characteristics. Similarly, for \acp{CrI}, using uninformative priors makes the interval estimation primarily driven by the annotation process -- that is, by the sample and its characteristics.  Instead, the main factor influencing the performance of the methods is the distribution of correctness labels across the dataset, or in other words, the accuracy level of the dataset. Consequently, the performance observed in large-scale datasets mirror those of small ones, motivating the consistency in the results of SYN 100M (large-scale) and YAGO, NELL, DBPEDIA, and FACTBENCH (small-scale)

As such, \ac{aHPD} proves again to be the most \rttmeta{statistically} efficient solution when the \ac{KG} accuracy is skewed (i.e., $\mu \in \{0.9, 0.1\}$) and a competitive approach in the symmetric case (i.e., $\mu=0.5$), where it achieves top performance alongside Wilson. This further consolidates \ac{aHPD} as the preferred solution in real-world applications.

It is worth mentioning that, with unbiased sampling strategies, if two datasets have correctness labels distributed in opposite proportions -- i.e., presenting symmetric accuracy values of $\mu$ and $1-\mu$, respectively -- the evaluation process incurs the same costs. In other words, the iterative procedure requires (approximately) the same number of triples to converge, regardless of whether the dataset accuracy is $\mu$ or $1-\mu$. This property explains the consistent performance of the methods on SYN 100M with $\mu=0.9$ and $\mu=0.1$.

\subsection{Robustness}
\label{subsec:robustness}

\begin{figure}[t!]
\centering
\includegraphics[width=\linewidth]{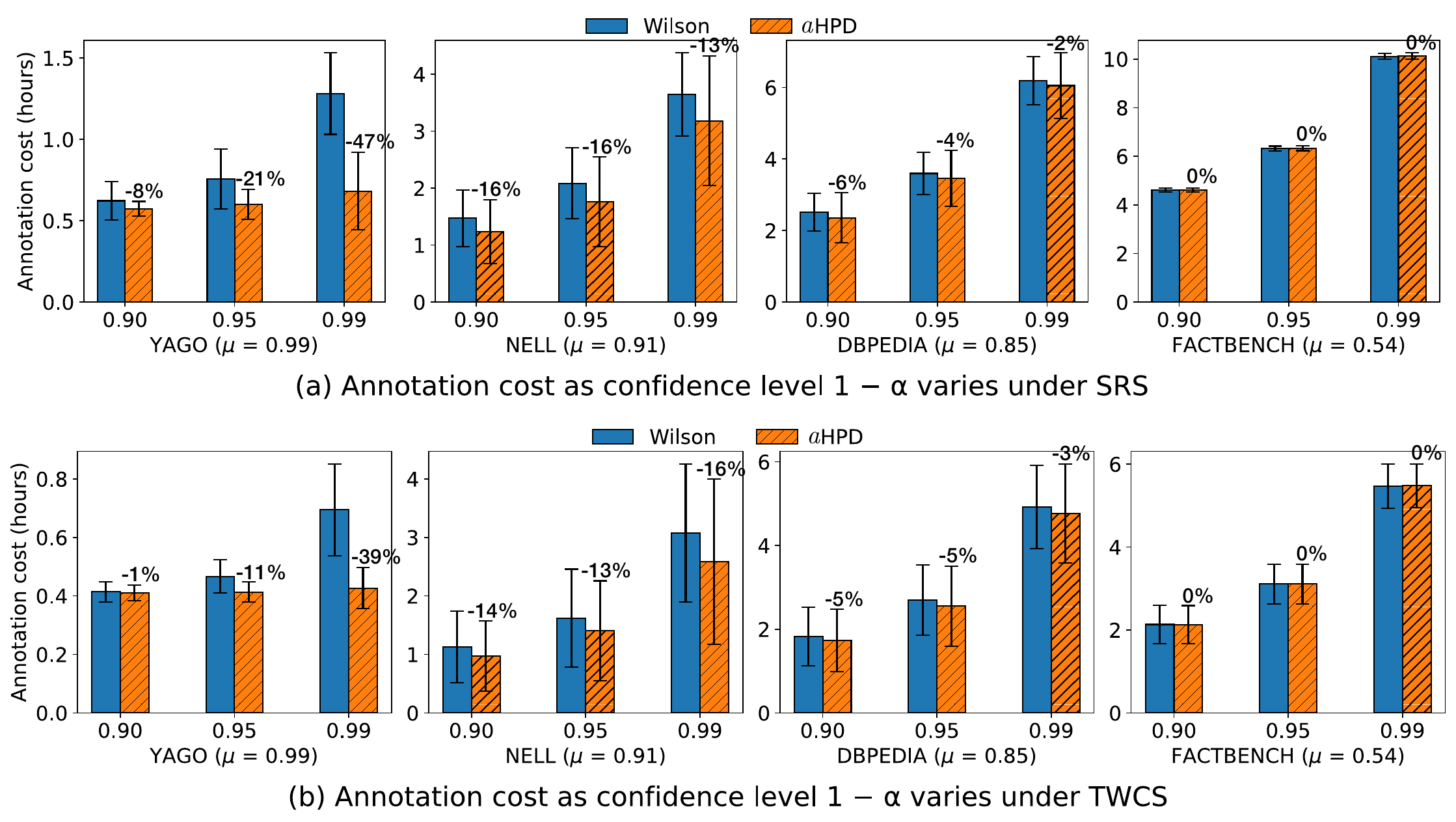}
  \caption{Annotation cost comparison between \textit{a}HPD and Wilson at different confidence levels $1-\alpha$ under SRS (a) and TWCS (b) on YAGO, NELL, DBPEDIA, and FACTBENCH KGs. We also report the reduction ratio (in \%) of \textit{a}HPD over Wilson.}
  \label{fig:robustness}
\end{figure}

Figure~\ref{fig:robustness} presents the annotation costs of \ac{aHPD} at various precision levels, comparing them to Wilson costs under \ac{SRS} and \ac{TWCS} sampling strategies. The reduction ratio of \ac{aHPD} relative to Wilson is also shown for each case. Since we have already demonstrated that \ac{aHPD} is a superior alternative to Wald, this comparison focuses solely on Wilson, the current state-of-the-art for \ac{KG} accuracy estimation~\cite{marchesin_silvello-2024}. 

The results in Figure~\ref{fig:robustness} confirm the superior performance of \ac{aHPD} not only at the standard $\alpha=0.05$ level but also at less ($\alpha=0.10$) and more ($\alpha=0.01$) stringent precision levels. Overall, \ac{aHPD} exhibits reduction ratios across all skewed datasets and precision levels, underscoring its versatility across diverse application scenarios. For instance, on YAGO, the cost reduction achieved by \ac{aHPD} reaches up to $47\%$ under \ac{SRS} and $39\%$ under \ac{TWCS} when $\alpha=0.01$. \rttmeta{To reinforce these findings, we conducted statistical analyses for $1-\alpha$ intervals with precision levels $\alpha \in \{0.1, 0.01\}$, obtaining results consistent with those found for $\alpha=0.05$ (cf. Table~\ref{tab:efficiency}). These outcomes further substantiate the higher efficiency of \ac{aHPD} across precision levels.}

The $\alpha=0.01$ setting is particularly noteworthy, as it reflects a high-precision scenario where a substantial degree of confidence in the estimates is essential. Such scenarios are common in fields like healthcare, where ensuring data quality is critical to prevent introducing inference errors into downstream applications that rely on the \ac{KG}. In this context, where the number of required annotations increases due to the stringent precision demands, the cost savings achieved by \ac{aHPD} become even more impactful. Since manual annotations in these cases often require expert knowledge, the ability to reduce costs while maintaining reliability is precious.

\rttmeta{Furthermore, it is also important to consider that, in real-world scenarios, the evaluation process typically involves multiple annotators (often 3-5) per fact, whose annotations are aggregated to determine the final correctness label. As the number of annotators increases, so do the corresponding costs. Depending on the available annotation budget, the cost reduction introduced by \ac{aHPD} can make the difference between an evaluation process that concludes successfully (due to convergence) and one that terminates prematurely (due to budget exhaustion).}

As expected, there are no benefits in using \ac{aHPD} over Wilson on FACTBENCH (quasi-symmetric) -- but no drawbacks either. This favors \ac{aHPD}, as it reduces annotation costs in skewed cases without introducing additional costs in quasi-symmetric ones.
Thus, the robustness analysis across different precision levels confirms the all-around superiority of \ac{aHPD} compared to the considered baselines, providing strong evidence to support its adoption with any sampling strategy and in any scenario where there is a need to evaluate \ac{KG} accuracy with limited annotations.

\section{Related Work}
\label{sec:relatedWork}

The efficient evaluation of \acp{KG} accuracy has been largely overlooked. The first approach, KGEval \cite{ojha_talukdar-2017}, is an iterative algorithm that alternates between control and inference stages. In the control stage, KGEval uses crowdsourcing to select facts for evaluation. In the inference stage, it applies type consistency and Horn-clause coupling constraints~\cite{mitchell_etal-2018,lao_etal-2011} to the evaluated facts to automatically estimate the correctness of additional facts. This process repeats until no more facts are evaluated. KGEval does not scale to real-life \acp{KG}~\cite{gao_etal-2019} and is susceptible to error propagation due to its probabilistic inference mechanism, making it unsuitable for our work.

To overcome KGEval limitations, \citet{gao_etal-2019} resorted to sampling strategies and estimators that gauge \ac{KG} accuracy with statistical guarantees. To minimize annotation costs, \cite{gao_etal-2019} proposed the use of cluster sampling techniques (i.e., \ac{TWCS}) over standard sampling (i.e., \ac{SRS}). On the other hand, the authors relied on the Wald interval~\cite{casella_berger-2002}, known to have reliability issues when used on binomial proportions~\cite{brown_etal-2001,wallis-2013}. To address Wald issues for \ac{KG} accuracy evaluation, \citet{marchesin_silvello-2024} proposed to use the Wilson interval~\cite{wilson-1927}, being it better suited for binomial proportions.

Although reliable, the Wilson interval shares Wald's frequentist interpretation, preventing a one-shot probabilistic understanding of its reliability. Additionally, the interval balances efficiency and reliability, sometimes requiring more annotations than Wald for greater reliability. In contrast, our solution offers a one-shot interpretation of interval confidence, making it better suited for the task while remaining the most efficient option. Moreover, the proposed approach, $a$HPD, eliminates the need for analysts to select a specific prior, addressing a key barrier to using Bayesian methods.

\citet{qi_etal-2022} developed an efficient human-machine collaborative framework to minimize annotation costs through inference. This framework leverages statistical techniques similar to those in~\cite{gao_etal-2019} and combines inference mechanisms akin to those in~\cite{ojha_talukdar-2017}. Their work focuses on the interplay between human annotations and inference mechanisms rather than on \acp{CI} and their impact on the minimization problem. Therefore, we do not include their method in our experimental analysis, but the $a$HPD method can be integrated into \citet{qi_etal-2022}'s framework to enhance efficiency.

\section{Conclusions}
\label{sec:conclusions}

In this paper, we exposed the limitations of \acfp{CI}, used in current state-of-the-art solutions for evaluating \ac{KG} accuracy. Rooted in frequentist statistics, these efficient intervals often lead to interpretation fallacies regarding confidence, precision, and likelihood, making them inconvenient for the task. 
We defined the interval estimation problem in Bayesian terms to address these limitations, introducing Bayesian \acfp{CrI}. Using a prior distribution reflecting the initial belief about the \ac{KG} accuracy and updating it with observed data through the likelihood, \acp{CrI} define the range of probable parameters in the posterior distribution. In other words, these intervals represent where the \ac{KG} accuracy is likely to fall with $1-\alpha$ probability. Since they are built directly on the posterior distribution, \acp{CrI} avoid the interpretative pitfalls of \acp{CI}, offering reliable and efficient solutions in a \emph{one-shot} setting.

We considered two families of \acp{CrI}: \acf{ET} and \acf{HPD} intervals. Through theoretical and empirical analyses, we demonstrated that \ac{HPD} \acp{CrI} represent the most efficient solution in real-life cases -- where the \ac{KG} accuracy is skewed -- while remaining competitive with state-of-the-art in controlled scenarios -- where the \ac{KG} accuracy is (quasi-)symmetric.

We also addressed one of the major resistances to using Bayesian methods: the challenge of choosing the right prior. Prior selection is a critical aspect of Bayesian approaches, as selecting a poor prior can lead to deceptive interpretations or inefficient solutions. We proposed the \acf{aHPD} algorithm, where multiple objective priors are concurrently used to generate different $1-\alpha$ \ac{HPD} intervals. These intervals compete to achieve convergence in the minimization problem, ensuring the most efficient outcome is always obtained from the set of competing solutions. In this way, \ac{aHPD} removes the need to choose a specific prior in advance.

Through extensive experiments on both real and synthetic data, we showed that \ac{aHPD} outperforms the state-of-the-art approaches across different scales, providing robust results regardless of the precision required for the evaluation process. Notably, \ac{aHPD} reduced annotation costs by up to $47\%$ in high-precision scenarios. Based on this thorough evaluation, which involved both \ac{SRS} and \ac{TWCS} sampling strategies, we recommend that practitioners adopt \ac{aHPD} with \ac{TWCS} as the most efficient and reliable solution for auditing \ac{KG} accuracy in a one-shot setting.

\paragraph{\textbf{\rtwo{Future Work and Limitations}}} We plan to explore the use of the \ac{aHPD} algorithm in dynamic environments, where \ac{KG} contents evolve over time~\cite{polleres_etal-2023}. \rtwo{Consider a typical scenario for \acp{KG}, where batches of new content arrive at irregular intervals. Assume an initial evaluation has been conducted, resulting in a given \ac{KG} accuracy estimate. When the volume of new content reaches a certain, predefined mass, we can perform a new evaluation. Leveraging the Bayesian framework we introduced to define the problem, we can consider the initial estimate as prior knowledge and use it as an informative prior for the new evaluation process. Equipped with reliable informative priors, \ac{aHPD} can converge significantly faster than solutions based on uninformative priors (see Example~\ref{ex:infPrior}).}

\rtwo{Despite promising, this approach also hides potential limitations. Massive updates (e.g., half the size of the \ac{KG}) with significantly different accuracy levels can render the informative prior less effective and, potentially, even deceptive. Exploring these challenges represents a critical direction for future research.}

\begin{acks}
The work was supported by the HEREDITARY project, as part of the
EU Horizon Europe program under Grant Agreement 101137074. ChatGPT and Grammarly were used as writing assistants to help rephrase certain sentences in this manuscript.
\end{acks}

\bibliographystyle{ACM-Reference-Format}
\bibliography{biblio}

\end{document}